\documentclass[a4paper,11pt,headings=big,DIV=12]{scrartcl}
\pdfoutput=1 
\usepackage{amsmath,amssymb,mathtools}
\usepackage[colorlinks,linkcolor=mygray,urlcolor=blue,citecolor=blue]{hyperref}
\usepackage{tcolorbox} 
\definecolor{mygray}{gray}{0.2}
 \addtokomafont{disposition}{\rmfamily\boldmath}
  \usepackage{bbold}
 \renewcommand*{\arraystretch}{2.0}
\numberwithin{equation}{section}
 
\definecolor{mypink1}{rgb}{0.9, 0.2, 0.6}
\usepackage{graphicx}
\addtokomafont{disposition}{\rmfamily\boldmath}
\usepackage[utf8]{inputenc}
\usepackage{enumerate} 
\usepackage{slashed}
\usepackage{cite} 
\usepackage{comment}
\usepackage{multirow}
\usepackage{acronym}
 \usepackage{bbold}
\allowdisplaybreaks
\newcommand{\Ima}{\textrm{Im}}

\newcommand{\vev}[1]{\langle #1 \rangle} 
\newcommand{\state}[1]{|#1\rangle}

\newcommand{\matel}[3]{\langle #1|#2|#3\rangle}

\newcommand{\al}{\alpha}
\newcommand{\be}{\beta}
\newcommand{\ga}{\gamma}
\newcommand{\Ga}{\Gamma}
\newcommand{\de}{\delta}
\newcommand{\De}{\Delta}
\newcommand{\la}{\lambda}
\newcommand{\La}{\Lambda}

\newcommand{\sig}{ \sigma}

\newcommand{\TeV}{\,\mbox{TeV}}
\newcommand{\GeV}{\,\mbox{GeV}}
\newcommand{\MeV}{\,\mbox{MeV}}

\newcommand{\keV}{\,\mbox{keV}}

\newcommand{\LaQCD}{\Lambda_{\textrm{QCD}}}
\newcommand{\LaQCDIR}{\la_{\text{QCD}}}

\newcommand{\pl}{\!+\!}

\newcommand{\EQ}{Eq.~}
\newcommand{\EQs}{Eqs.~} 
\newcommand{\TAB}{Tab.~}

\newcommand{\FIG}{Fig.~}

\newcommand{\SEC}{Sec.~}
\newcommand{\SECs}{Secs.~}
\newcommand{\APP}{App.~}
\newcommand{\APPs}{Apps.~}
\newcommand{\REF}{Ref.~}

\newcommand{\Dla} {\overset{\leftarrow}{D}}
\newcommand{\Dra} {\overset{\rightarrow}{D}}

\newcommand{\Lag}{{\cal L}}
\newcommand{\LagImp}{{\cal L}^R}

\newcommand{\dphi}{d_\varphi}

\newcommand{\NNone}{{\cal N}=1}
\newcommand{\LO}{{\text{LO}}} 
\newcommand{\NLO}{{\text{NLO}}} 
\newcommand{\dxchiPT}{{(d)$\chi$PT }}
\newcommand{\chiPT}{{$\chi$PT} }
\newcommand{\dchiPT}{{d$\chi$PT} }
\newcommand{\LNc}{large-$\Nc$ }

\newcommand{\Lageff}{\Lag_{\text{eff}} }
\newcommand{\LagLOchiPT}{\Lag^{\chi\text{PT}}_{\text{LO}} }
\newcommand{\LagLOdchiPT}{\Lag^{\text{d}\chi\text{PT}}_{\text{LO}} }
\newcommand{\LagLO}{\Lag_{\text{LO}} }

\newcommand{\Lagmq}{\Lag_{m_q}}

\newcommand{\Lagkinpi}{\Lag^{(\pi)}_{\text{kin}}}
\newcommand{\LagkinD}{\Lag^{(D)}_{\text{kin}}}

\newcommand{\Laganom}{\Lag_{\text{anom}}}

\newcommand{\XX}{\chi}
\newcommand{\XXh}{\hat{\XX}}

\newcommand{\el}{\text{el}}
\newcommand{\ma}{\text{mag}}
\newcommand{\gast}{\ga_*}
\newcommand{\best}{\be_*}
\newcommand{\gastp}{\ga'_*}

\newcommand{\bestp}{\be'_*}
\newcommand{\bestpp}{\be''_*}
\newcommand{\gaG}{\ga_{G^2}}
\newcommand{\gastG}{(\ga_{G^2})_*}
\newcommand{\gm}{g}
\newcommand{\Deqq}{{\De_{\bar qq}}}

\newcommand{\qq}{\vev{\bar qq}}

\newcommand{\QCD}{\text{QCD}}
\newcommand{\CDQCD}{\text{CDQCD}}

\newcommand{\CFT}{\text{CFT}}

\newcommand{\TT}{\TEMT}
\newcommand{\OSa}{{S^a}}

\newcommand{\OPa}{{P^a}}

\newcommand{\OP}{{P}}
\newcommand{\OS}{{S}}

\newcommand{\Tr}{\mathrm{Tr}}
\newcommand{\GaF}[1]{\bar{\Ga}_{#1}}
\newcommand{\GaE}[1]{\Ga_{#1}}
\newcommand{\CoE}[1]{\vev{ {#1}(x){#1}(0)} }

\newcommand{\ORD}{{\cal O}}

\newcommand{\dil}{D}

\newcommand{\IRstate}{\varphi}

\newcommand{\mq}{m_q}
\newcommand{\dbar}{{d-1}}
\newcommand{\TEMT}{ \Tud{\rho}{\rho} }

\newcommand{\Tud}[2]{T^{#1}_{\;\; #2}}

\newcommand{\mink}{\eta}

\newcommand{\sigH}{D}

\newcommand{\weyl}{s}

\newcommand{\Oone}{{\cal O}_1}
\newcommand{\Otwo}{{\cal O}_2}

\newcommand{\Nf}{N_f}
\newcommand{\Nc}{N_c}
\newcommand{\Op}{{\cal O}}
\newcommand{\Opqq}{{\cal O}_{\bar qq}}

\newcommand{\mbase}{\overline{m}_{D}}

\definecolor{violet}{rgb}{0.94, 0.2, 0.8}
\definecolor{lightblue}{rgb}{0.39, 0.58, 0.93} 
\definecolor{lightgreen}{rgb}{0.1, 0.73, 0.33}
\newcommand{\comb}[1]{ 
{#1}}

\DeclareOldFontCommand{\tt}{\normalfont\ttfamily}{\mathtt}

\usepackage[titles]{tocloft}

\begin{document}

\begin{flushright}
CERN-TH-2023-201\\
\today
\end{flushright}

\vspace{3mm}

\begin{center}
{\Large\bfseries \boldmath QCD with an Infrared Fixed Point and a Dilaton }
\\[0.8 cm]
{\large%
  Roman Zwicky$^{a,b}$
\\[0.5 cm]
\small 
 $^a$ Higgs Centre for Theoretical Physics, The University of Edinburgh, \\
Peter Guthrie Tait Road, Edinburgh EH9 3FD, Scotland, UK \\[0.2cm]
$^b$ Theoretical Physics Department, CERN, \\
Esplanade des Particules 1,  Geneva CH-1211, Switzerland
} \\[0.5 cm]
\small
E-Mail:
\texttt{
\href{mailto:Roman.Zwicky@ed.ac.uk}{Roman.Zwicky@ed.ac.uk}}
\end{center}

\bigskip
\thispagestyle{empty}

\begin{abstract}\noindent 
Following our previous work, we further investigate the possibility that the chirally broken phase of gauge theories admits an infrared fixed-point interpretation. 
The slope of the  $\beta$ function, $\beta'_*$, is found to 
vanish at the infrared fixed point which has several attractive features,  including logarithmic running.  
We provide a more in-depth analysis of our previous result  
that the mass anomalous dimension assumes $\gamma_* = 1$ at the fixed point. 
The results are found to be 
consistent  with  ${\cal N}=1$ supersymmetric gauge theories.
In a second part the specific properties of a dilaton, the (pseudo) Goldstone 
due to spontaneous symmetry breaking, are investigated.  
Dilaton soft theorems indicate that a soft dilaton mass can only originate from an operator of scaling dimension two. 
In the gauge theory  this role is taken on by the  $\bar qq$-operator.
The QCD dilaton candidate,  the $\sigma = f_0(500)$ meson, is  investigated 
and  singlet-octet mixing is found to be relevant.    
Finally, we briefly discuss the dilaton as a potential Higgs boson candidate, which requires the ratio of dilaton to pion decay constants to be close to unity.
In QCD this condition is approximately satisfied, though it remains unclear whether this is accidental or a consequence of a yet to be uncovered principle.
 \end{abstract}

\mdseries 

\tableofcontents
\thispagestyle{empty}


\pagestyle{plain}

\section{Introduction}
\label{sec:intro}

The idea that the strong interaction can be described by an infrared (IR) fixed  point  
predates QCD itself \cite{Isham:1970xz,Isham:1970gz,
Ellis:1970yd,Ellis:1971sa,Crewther:1970gx,Crewther:1971bt}.  
In our previous work \cite{Zwicky:2023bzk}, it was found 
in three different ways, that pion physics can be reproduced when the 
quark-mass anomalous dimension at the IR fixed point assumes $\gast = 1$. 
The methods included  the hyperscaling relation for the pion mass, compatibility of  
the Feynman-Hellmann theorem with the  trace of the energy momentum tensor (TEMT),  
and matching a long distance correlator. 
The main idea that underlies the scenario is that the gauge theory flows into an IR fixed point and that it is 
the quark condensate $\vev{\bar qq} \neq 0$ that spontaneously breaks
breaks \emph{both}  chiral and  
conformal symmetry.\footnote{Throughout this paper we do not distinguish 
conformal and scale symmetry. In $d=4$ it is believed 
that these are equivalent for non-trivial unitary theories
(cf. the review \cite{Nakayama:2013is}  or  \cite{Zwicky:2023bzk} for more comments and references).} 
This leads, besides the pion, to one additional (pseudo) 
Goldstone \cite{Isham:1970gz,Coleman:1985rnk,Low:2001bw}, 
the dilaton.  The fact that the conformal symmetry is only emergent and that the dilaton has 
vacuum quantum numbers complicates  matters. 
Let us turn to the assumptions. The minimal implementation of IR conformality, adopted in \cite{Zwicky:2023bzk}, is that  the TEMT on single particle IR-states $\IRstate$, which includes the vacuum, the pion and possibly the dilaton 
\begin{equation}
\label{eq:IRCFT1}
 \matel{\IRstate'(p')}{\TEMT}{\IRstate(p)} _{q =0}  =  0 \;,  
\end{equation}
vanishes for zero momentum transfer $q = p - p'$. 
 In a true CFT \eqref{eq:IRCFT1} holds on all physical states of all momentum transfers, and can be seen as its definition.\footnote{\label{foot:mN} In the presence of a massless dilaton \eqref{eq:IRCFT1} holds for the LO EFT: 
$\TEMT|_{\LO} =0$ \cite{Zwicky:2023fay}; and  extends to the case of  single nucleon states 
$\matel{N(p')}{\TEMT}{N(p)} _{q =0}  =  0$  despite $m_N \neq 0$ \cite{DelDebbio:2021xwu,Zwicky:2023fay}.} 
We make the reasonable assumption that if \eqref{eq:IRCFT1} holds true that there 
exists a scheme with vanishing  $\be$-function in the IR: $\best \equiv \be|_{\mu=0}=0$.\footnote{
{Note that $\be(g^*)=0$  is invariant under analytic redefinitions of $g$ 
\cite{Collins:1984xc} but not necessarily when non-analytic (e.g. canonical versus holomorphic coupling 
in ${\cal N}=1$ supersymmetric QCD \cite{Shifman:1986zi,Arkani-Hamed:1997qui,Terning:1997xy}). 
In such exotic schemes the physics is hidden away in the field strength since it is the product 
 $\TEMT = \frac{\be(\mu)}{2 g(\mu)} G^2(\mu)$ which is physical and not the $\be$-function itself. 
 The EFT formulation is valid in schemes where $\be(g^*)=0$ only. }}  
Additionally, we assume that correlators obey conformal field theory (CFT) scaling in the deep-IR
\begin{equation}
\label{eq:scale1}
\vev{\Op(x) \Op(0)} \propto \frac{1}{(x^2)^{\De_{\Op}}}  + \text{GB-corrections} \;,
\end{equation}
corrected by terms due to Goldstone boson (GB). Above $\De_{\Op} = d_{\Op} + \ga_{\Op}$ is the scaling dimension,
 defined as the sum of the engineering  and the anomalous dimension. 
We will see that \eqref{eq:scale1}  essentially emerges by  combining  RG  and  EFT reasoning. 
Following the terminology in  \cite{DelDebbio:2021xwu} we will refer to this scenario as  conformal dilaton (CD) QCD.

It is widely believed that the dilaton retains a \emph{chiral mass} (no explicit symmetry breaking) 
\begin{equation}
\label{eq:base}
\mbase \equiv m_D|_{\mq=0} \;,
\end{equation}
in the case of an IR-emergent conformal symmetry.\footnote{Throughout this work we refer 
to the dilaton as $D$ when generic matters are discussed and to $\sig  (\leftrightarrow f_0(500))$ when referring to the dilaton candidate in QCD and to $h$ in the context of the Higgs boson.}
\setcounter{page}{1}
To the best of our knowledge this issue has never been fully settled.\footnote{In the context of gauge theories, it is generally believed that  $\mbase$ becomes larger with respect to the 
other hadronic scales as one moves away from the conformal window by lowering the number of flavours.}
We  take a pragmatic attitude by studying both cases separately.  {(We stress that all other hadrons remain massive 
in the chiral limit, primarily since there is no dynamical reason for them to be massless cf. footnote \ref{foot:mN}.)}
For  $\mbase = 0$ we use dilaton-\chiPT (\dchiPT\!\!) and 
for $\mbase \neq 0$ we use standard chiral perturbation theory (\chiPT\!\!) since the dilaton can be integrated out. 
Phenomenologically a dilaton approach might still be of interest if $\mbase \ll  \La_{\text{hadron}} = \ORD(m_N)$ which is what 
most approaches have in mind.   
This includes lattice Monte Carlo investigation of gauge theories  \cite{LatticeStrongDynamics:2023bqp,Hasenfratz:2020ess,Kuti:2022ldb,Fodor:2019ypi,Fodor:2017gtj,Fodor:2018uih,Fodor:2019vmw,Fodor:2020niv,Chiu:2018edw,LatticeStrongDynamics:2018hun,LatKMI:2016xxi,DelDebbio:2015byq,Brower:2015owo,LSD:2014nmn,LatKMI:2014xoh},  
EFT descriptions thereof \cite{Golterman:2016lsd,Appelquist:2017wcg,Appelquist:2022mjb,Freeman:2023ket};
the dilaton as a Higgs boson within \cite{Matsuzaki:2012xx,Dietrich:2005jn,Cata:2018wzl} and investigations not necessarily  linked to gauge theories \cite{Goldberger:2007zk,Appelquist:2010gy,Bellazzini:2012vz,Chacko:2012vm,Coradeschi:2013gda}. 
Or the dilaton as a driving force of inflation \cite{Csaki:2014bua}, in dense nuclear interactions   \cite{Ma:2019ery,Rho:2021zwm,Brown:1991kk}, 
in cosmology \cite{Wetterich:1987fm,Shaposhnikov:2008xb,Shaposhnikov:2008xi} and
   the $\sig$-meson in QCD \cite{Crewther:2012wd,Crewther:2015dpa}. 
 The name dilaton is also used in composite models for a light $0^{++}$-partner of pseudo Goldstones  
    \cite{Bruggisser:2022rdm}.

Let us briefly summarise the main findings of our work. By RG methods it is inferred that 
the slope of the $\be$-function vanishes  at the fixed point: $\bestp=0$.  A result with many attractive features,  
as it is compatible with ${\cal N}=1$ supersymmetric gauge theories, makes a light dilaton more likely
and  implies logarithmic running near the fixed point as one would expect from the EFT itself.  
A more thorough justification  for $\gast =1$ in the context of matching correlation function is given 
combining  RG  and  EFT methods.  Applying a double-soft dilaton theorem, it is found that 
an operator $\Op$ giving a mass  to the dilaton must be of scaling dimension $\De_{\Op}=2$. The second part of the paper is more 
qualitative and contains a discussion on whether a dilaton can  be massless, applies \dchiPT 
to the $\sig$-meson in QCD, and concerns the dilaton as a Higgs boson. 
The main outstanding questions are (i) the size of the chiral mass $\mbase$ (as the distance from the conformal window)
and (ii)   the ratio of decay constants  $r_{N_f} = {F_\pi}/{F_D}$ as a function of the number of flavours $N_f$. 
If $r_2 \approx 1$ holds,  this would provide the rationale for coupling the dilaton like a Higgs and 
quite might make it compatible with Large Hadron Collider (LHC) constraints.

The paper is organised as follows. 
In \SEC\ref{sec:QCDEFT} the gauge theory and  \dchiPT are defined including a review of the  conformal window. 
In \SEC\ref{sec:deep}  QCD correlators   matched to the EFTs  in the deep-IR and 
consistency with   $\NNone$ supersymmetric  gauge theories is discussed in \SEC\ref{sec:SUSY}.  
In  \SEC\ref{sec:soft} soft and double-soft dilaton theorems are exploited \SEC\ref{sec:soft}. 
The remaining part consists of a discussion of mass or no mass for the dilaton 

Arguments for and against  a massless dilaton are addressed in \SEC\ref{sec:massless}.  
The dilaton as the $\sig$-meson in QCD and  as the Higgs boson are discussed in \SECs  \ref{sec:QCD}  
and \SEC\ref{sec:outlook} respectively. 
The paper ends with  a summary and conclusions  in \SEC\ref{sec:conc}.  Conventions,
 more on soft theorems  and details on mixing   are deferred to \APPs \ref{app:conv}, \ref{app:soft}
and \ref{app:mix} respectively.

\section{The Gauge Theory and its Low Energy Effective Theory }
\label{sec:QCDEFT}

The core ingredients to this work are the gauge theory, its RG quantities, and the low energy EFT which 
corresponds to \chiPT and \dchiPT when a dilaton is added and are 
 introduced  in minimal form below.  

\subsection{The gauge theory  and its renormalisation group quantities}

The gauge theory Lagrangian is given by 
\begin{equation}
\label{eq:LQCD}
{\cal L}_{QCD} = - \frac{1}{4}G^2  + \sum_{q=1}^{\Nf} \bar q (i \slashed{D} -\mq) q \;,
\end{equation}
where  $G^2 = G_{\mu\nu}^a G^{a\, \mu\nu}$ is the field strength tensor with $a$ denoting the adjoint index of the 
 gauge group, $D_\mu = (\partial + i g A)_\mu$  is the gauge-covariant derivative and the quarks are in some 
 unspecified representation of the gauge group. For the more formal part of the paper the $\Nf$ flavours are assumed to be degenerate.
There are two parameters in the Lagrangian, the gauge coupling  $g$ and the quark mass $\mq$,  
whose RG-behaviour we must track. 
The gauge coupling 
is relevant and irrelevant in the UV and IR respectively whereas for the quark mass it is the other way around.  
Global flavour symmetries are discussed in the next section. 
 
The anomalous dimensions of the parameters  and their conjugate operators are defined from the
renormalised quantities as follows
\begin{alignat}{4}
\label{eq:ga}
&  {\be} &\;\equiv\;&  \phantom{-} \frac{d}{d\ln \mu}  g   \;,  \qquad & &   \gaG &\;\equiv\;&  - \frac{d}{d \ln \mu} \ln G^2  \;,   \nonumber \\[0.1cm]
&   \ga_m   &\;\equiv\;& - \frac{d}{d\ln \mu} \ln \mq      \;, \qquad & &      \ga_{\bar qq}  &\;\equiv\;&  - \frac{d}{d\ln \mu} \ln  \bar qq     \;.
\end{alignat} 
A central quantity to this work is the TEMT \cite{Crewther:1972kn,Chanowitz:1972vd,Chanowitz:1972da,Minkowski:1976en,Adler:1976zt,Nielsen:1977sy,Collins:1976yq}  
 \begin{equation}
 \label{eq:TEMTphys}
 \TEMT|_{\text{phys}} =  \frac{\be}{2 g} G^2 + \sum_q \mq(1\pl \ga_m)    \bar{q}q  \;,
 \end{equation}
 since its vanishing on physical states signals conformality.  
 The subscript ``phys" indicates that we have omitted terms  proportional to 
the equation of motion (and BRST exact terms which arise upon gauge fixing)  
which, nota bene, vanish on physical states \cite{Adler:1976zt,Nielsen:1977sy,Collins:1976yq} . It is very useful to note that \eqref{eq:TEMTphys} consists of two 
separately RG-invariant terms $\TEMT = \Oone + \Otwo$  
\begin{equation}
 \label{eq:inv1} 
 \Oone =  \frac{\be}{2g}G^2+ \sum_q  \ga_m  \mq \bar qq    \;, \quad   \Otwo = \sum_q  \mq \bar qq\;.
\end{equation}
Imposing $\frac{d}{d\ln \mu} {\cal O}_{1,2} =0$ results in 
\begin{equation}
\label{eq:def}
\gaG 
= \be' - \frac{\be}{g} \;, \quad \ga_{\bar qq} = - \ga_m \;,
\end{equation}
if one considers a quark-mass-independent scheme for the $\be$-function.  
Quantities at the IR fixed point are denoted by a star 
\begin{equation}
\label{eq:gast}
\gastG  =  \bestp \;,           \quad \gast \equiv  (\ga_m)_*\; ,
\end{equation}
where $\best =0$ has been assumed as stated earlier. In the context of conformal field theories (CFT) e.g  
\cite{Mack:1988nf,DiFrancesco:1997nk,Braun:2003rp,HOCFT,Rychkov:2016iqz,Poland:2018epd},  
 the important quantities are the operator scaling dimensions $\De_\Op \equiv d_\Op + \ga_\Op$ (sum of engineering  the anomalous dimension)
\begin{alignat}{3}
\label{eq:De}
& \De_{G^2}  &\;=\;& d + \bestp  &\; \to  \;& \quad 4 + \bestp \;, \nonumber \\[0.1cm] 
& \De_{\bar qq}  &\;=\;& (d-1) - \gast  &\; \to  \;& \quad 3  - \gast \;.
\end{alignat}
The  $d$-dimensional expressions have been given for later use.  

\subsection{Synopsis of the conformal window}
\label{sec:CW}

\begin{figure}[h!]
 \begin{center}
\includegraphics[width=0.8\linewidth]{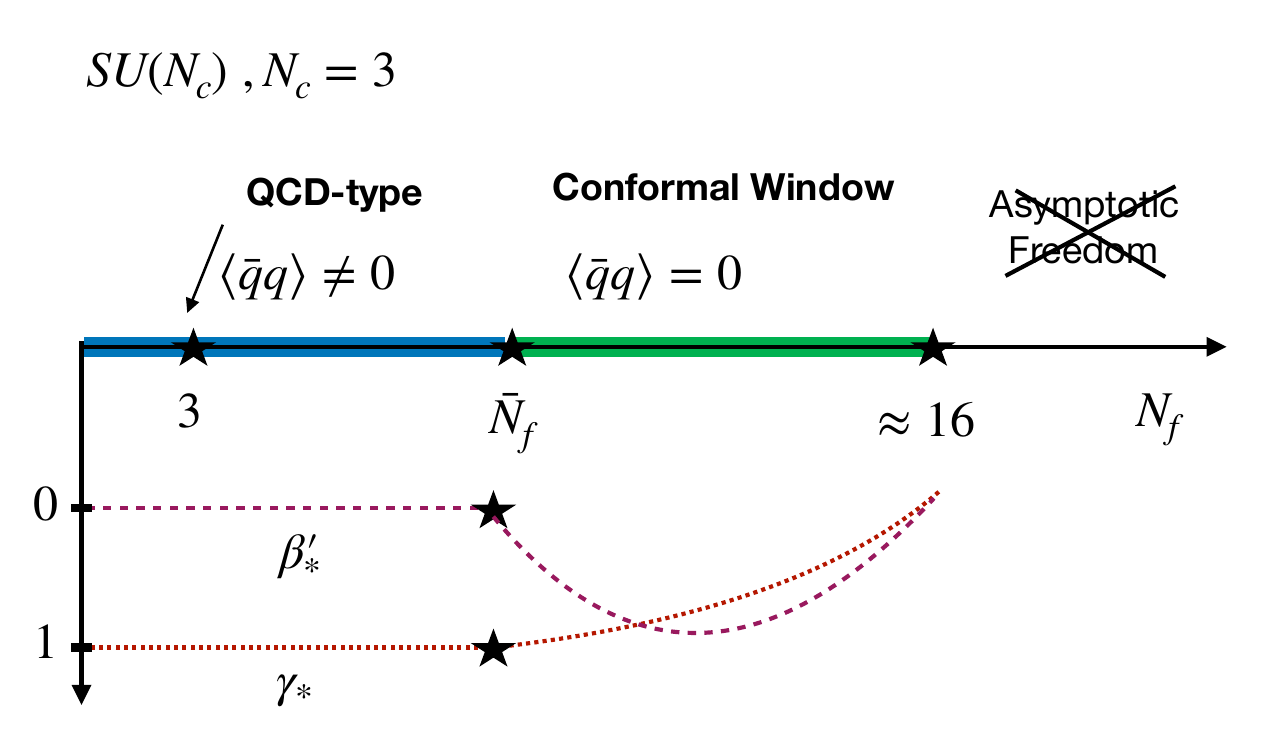}
\vspace{-0.5cm}
	\caption{\small Sketch of the conformal window for $G = SU(N_c)$ with $N_c =3$. For $N_f > 16$ 
	asymptotic freedom is lost and those theories are not considered. Below there is the 
	perturbative Caswell-Banks-Zaks  IR fixed point. 
	As $\Nf$ is lowered the IR fixed point coupling becomes stronger until 
	chiral symmetry breaking sets in (for some unknown critical $\bar{N}_f$).  
	In this paper 
	we explore the possibility that the theory admits an IR fixed point interpretation in parts or all of 
	the broken phase as well. 
	The evolution of the IR fixed point quantities $\gast$ and $\bestp$, to be deduced in \SEC\ref{sec:deep}, 
	 are shown  as a function of $\Nf$. {We stress that the evolution between $\bar{N}_f$ and $16$ 
	 is only schematic (the question of continuity of the transition at $\bar{N}_f$ cannot be assessed since $N_f$ is a discrete number) 
	 and most importantly that the evolution below $\bar{N}_f$ is non-standard (not certain) but shown to be consistent with  the IRFP assumption. 
	 In  \SEC\ref{sec:SUSY} arguments are given in favour of this picture for ${\cal N}=1$ supersymmetric 
	 gauge theories.}   }
	\label{fig:CW}
	\end{center}
\end{figure}

For the reader not acquainted with the basic notation of the conformal window 
we briefly  summarise  its minimal content  (the ${\cal N}=1$ supersymmetric case is described in \SEC\ref{sec:SUSY}). 
The conformal window is the study of  different phases of gauge theories, cf. \FIG\ref{fig:CW},  
as a function of the the gauge group e.g. $SU(N_c)$ and the  quark representation of $\Nf$ 
massless fermions. The conformal window has first been intensely studied  within technicolor  
 model building  (reviewed in 
\cite{Hill:2002ap,Sannino:2009za,Cacciapaglia:2020kgq}). The current understanding is largely based  on the ${\cal N}=1$ supersymmetric case and non-supersymmetric lattice studies. 
The main point is that in the $(\Nf,\Nc)$-plane, in the domain of asymptotic freedom, 
there are two phases:  one where the theory flows into an IR fixed point, and the  QCD-phase.  
In the latter there is confinement and  chiral symmetry is spontaneously broken
 by the quark condensate $\vev{\bar qq} \neq 0$  
(the global flavour symmetry breaks from $SU(\Nf)_L \otimes SU(\Nf)_R \to  SU(\Nf)_I$ 
to the diagonal subgroup accompanied by $N_f^2-1$ massless pions).   
If we fix $N_c =3$ then we know that around $\Nf \approx 16$ the theory 
admits  a perturbative (Caswell-Banks-Zaks  \cite{Caswell:1974gg,Banks:1981nn}) IR fixed point 
 and that somewhere in 
between $N_f =16 $ and $\Nf=3$ (QCD) there is the transition into the QCD phase. 
For what critical $\bar{N}_f$ this happens  is a matter of intense debate,
 but known exactly in the ${\cal N}=1$  case.

\subsection{Low energy effective theory --  dilaton-\chiPT at LO}
\label{sec:EFT}

In this section  the dilaton EFT is discussed, its starting point is the well-understood  \chiPT \cite{Gasser:1983yg,Donoghue:1992dd,Leutwyler:1993iq,Scherer:2012xha} which is the EFT of 
$N_f^2-1$ pions describing a wealth of  low-energy data.
Based on this successful framework the  dilaton, the Goldstone due to the scale symmetry breaking, is added. 
The Goldstone bosons are parameterised in exponential form 
\begin{equation}
\label{eq:Uchi}
U = e^{ i2  \pi^a T^a /F_\pi} \;,\quad \hat{\chi} \equiv  e^{-\frac{D}{F_D}}  \quad 
\comb{( \chi  \equiv     F_D/\dphi  \hat{\chi}^{\dphi} \;, \quad \dphi \equiv \frac{d-2}{2} ) }  \;,
\end{equation} 
where  the quantities $F_{\pi}$ and $F_{D}$ are fundamental, in that they serve as  order 
parameters of the symmetry breaking 
\begin{equation}
\label{eq:fdecay}
  \matel{\pi^b(q)}{J_{5\mu}^a(x)}{0} = 
  i F_\pi  q_\mu  \de^{ab}  e^{iqx}  \;, \quad   
   \matel{\dil(q)}{J^{\dil}_\mu(x)}{0} =  i {F_{\dil}} q_\mu e^{iqx} \;,
   \end{equation}
defined as the matrix elements of the global symmetry currents 
\begin{equation}
J_{5\mu}^a(x) = \bar{q}(x) T^a \ga_\mu \ga_5 q(x) \;, \quad J^{\dil}_\mu(x) = x^\nu T_{\mu\nu}(x) \;,
\end{equation} 
between the vacuum and  the corresponding Goldstone. It is noted that the equation for the dilaton in \eqref{eq:fdecay} is equivalent to
\begin{equation}
\label{eq:FDTmunu}
\matel{\dil(q)}{T_{\mu\nu}}{0} =  
\frac{F_{\dil} }{\dbar}  
(m_{\dil}^2 \mink_{\mu\nu} - q_\mu q_\nu) \;,
\end{equation} 
when taking into account that $x^\nu \to  -i \partial_{q_\nu}$. 
From the zeroth component of the currents one can see heuristically that the charge of the symmetry does not annihilate the vacuum, thereby signalling spontaneous symmetry breaking (SSB).  The symmetries of the Goldstones are said to be realised non-linearly 
\begin{equation}
U \to  L U R^\dagger \;,  \qquad 
\comb{ \hat{\XX} \to \hat{\XX} e^{ \al(x)}  \quad  ({\XX} \to {\XX} e^{ \al(x)\dphi}) }  \;,
\end{equation}
where $(L,R) \in  SU(\Nf)_L \otimes SU(\Nf)_R$ and $\al(x) \in \mathbb{R}$ parameterise the standard chiral and  the Weyl  transformation respectively.  The latter can be interpreted as the implementation of scale transformations 
on the metric $\gm_{\mu\nu}$, rather than the coordinates,  and the fields $\varphi$
\begin{equation}
\label{eq:Weyl0}
\gm_{\mu\nu} \to e^{-2 \al(x)} \gm_{\mu\nu} \;, \quad \varphi \to e^{\weyl_\varphi \al(x)} \varphi \;,
\end{equation}
where $\weyl_\varphi$ is called the Weyl-weight which is not to be confused with the engineering 
dimension $d_{\varphi}$ of the field.  The Weyl-weight of the metric is $\weyl_{\gm_{\mu\nu}} = -2$ since 
it naturally contracts two coordinate vectors $x^2 = \gm_{\mu\nu} x^\mu x^\nu$ which would transform as 
$x^\mu \to e^{\al(x)} x^\mu$. 
Local Weyl invariance,  
\begin{equation}
\label{eq:Weyl-inv}
\sqrt{- \gm} \Lag \to ( e^{-\al d}\sqrt{- \gm} )  (e^{\al d}\Lag) = \sqrt{- \gm} \Lag \;,
\end{equation}
 is the guiding principle for the low energy EFT. The $\al$ space-dependence has and will be suppressed 
 occasionally  hereafter.  
 In the case of explicit and anomalous symmetry breaking the spurion technique is adapted. 
 
The  leading order (LO) Lagrangian consists of
\begin{equation}
\label{eq:LO0}
\LagLO  = \Lagkinpi  +  \LagkinD  + \LagImp_d  + \Lagmq +  \Laganom - V(\XX)   \;,
\end{equation}
two kinetic-, an improvement-, a quark mass-, an anomaly-  and a dilaton potential-term. 
The kinetic  terms  read
 \begin{equation}
 \label{eq:Lagkin}
 \Lagkinpi  =         \frac{ F_\pi^2 }{4}  \hat{\XX}^{d-2}  \Tr [ \partial^\mu  U \partial_\mu  U^{\dagger}]  \;, \quad 
 \LagkinD  =   \comb{ \frac{1}{2}   (\partial\XX)^2}  \;.
  \end{equation}
 The pion kinetic term is standard and the prefactor $\XX^2$ indicates that the pion has zero Weyl-weight which 
 can be deduced directly from the 
 conformal algebra  \cite{Ellis:1970yd}. 
\comb{Coupling the dilaton to the Ricci tensor
 \begin{equation} 
\label{eq:improve}
\LagImp_d = -   \frac{\dphi}{4(d-1)} \, R \, \XX^{d-2} \to -  \frac{1}{12} \, R\, \XX^2 \;,  \quad
\end{equation}
renders the dilaton kinetic term locally Weyl-invariant which 
has many advantages, see \REF\cite{Zwicky:2023fay}}.\footnote{The term 
\eqref{eq:improve} is the adaption of the improved EMT  \cite{Callan:1970ze} to the Goldstone case 
for which the presence of the dilaton is needed to improve the pion. 
\comb{ It leads to $\TEMT|^{\text{d}\chi\text{PT}}_{\LO} =0$ 
for $\mq=0$, and it improves flow theorems \cite{Zwicky:2023fay}.}}
 The quark-mass term reads  (${\cal M} \equiv \text{diag}(m_{q_1}, \dots , m_{q_{\Nf}})$)
\begin{equation}
\label{eq:Lagmq}
 \Lagmq  = 
\frac{ B_0 F_\pi^2}{2}  \Tr[{\cal M}U +U^\dagger {\cal M}^\dagger]  \hat{\XX}^{\De_{\bar qq}} 
 \;, \quad 
B_0 \equiv  -\frac{\vev{\bar qq}}{ F_\pi^2} \;,
\end{equation}
where $B_0$ assures that the Gell-Mann Oakes Renner (GMOR)-relation
$m_\pi^2 F_\pi^2 = - 2 m_q \qq$  \cite{Gell-Mann:1968hlm,Scherer:2012xha} 
 is reproduced. 
  The chiral symmetry is formally restored by assigning 
 the spurious transformation rule  ${\cal M} \to L^\dagger{\cal M}R$ which in turn dictates the form in   \eqref{eq:Lagmq}. 
 The quantities $B_0$ and $F_\pi$ are the two (independent)  low energy constants of LO $\chi$PT.   
 The quark  mass corresponds to explicit Weyl symmetry breaking but the latter can be restored by assigning the 
 spurious transformation 
$  \mq \to e^{(1+ \ga_m)\al} \mq$
and the 
  extra factor $\hat{\XX}^{\De_{\bar qq}}$ is then required to render $\sqrt{-\gm} \Lagmq $ Weyl invariant.  
  Alternatively, we may regard $m_q$ as the source for $\bar qq$ with the factor $\hat{\XX}^{\De_{\bar qq}}$ capturing 
  its scaling. 
 The remaining two terms  are more delicate and need more elaboration.

The anomalous term reads $\TEMT =\frac{\be}{2 g} G^2$ in the  $\mq \to 0$ limit. 
It is not well understood whether 
a term needs to be added for the trace anomaly in the EFT in analogy with the WZW-term for the chiral anomaly  in \chiPT 
(e.g.  \cite{Donoghue:1992dd}). Clearly, the leading effect would be captured by $\bestp$ and is sometimes parameterised in the 
 EFT as $\bestp$ times  operators in the LO TEMT cf. \cite{Crewther:2015dpa,Cata:2018wzl}
(and also  \cite{Bellazzini:2012vz}). 
Using RG methods we will establish $\bestp=0$  in  \SEC\ref{sec:deep} which  implies that  $\de g = g -g_*$  runs logarithmically instead of power-like close the the fixed point.  We would therefore think  that the EFT can capture these effects 
through its own loops and matching of NLO low energy constants with QCD. 
Hence, we will drop  $\Laganom $ from the LO Lagrangian.

We turn to the discussion of the potential.  
It has been recognised by Zumino a long time ago \cite{Zumino:1970tu}
  that, at least in the absence of anomalous breaking, the only permissible term in the potential is 
  $V_Z(\XX) = \la \XX^d$. 
 Such a term on its own is troublesome, cf. \REF \cite{Cata:2018wzl} for detailed discussion,
  as it generates  a linear as well as a mass term.  
 Hence it was concluded that $\la$ must be a function of a symmetry breaking parameter.  
 A simple and frequently used form of a potential is given 
 by adding another power $\XX^\De$ which can be written as\footnote{In the literature this potential is often discussed 
 without specific reference to the $\bar qq$-operator. For  $\De \to 4$ it assumes the famous logarithmic form  
 used as an ansatz for an EFT of pure Yang-Mills \cite{Migdal:1982jp}.}
\begin{alignat}{2}
\label{eq:V}
&  V_{\De} (\hat{\chi}) &\;=\;&   \frac{m_D^2 F_D^2 }{ \De -d} \left( \frac{1}{\De} \hat{\chi}^\De - \frac{1}{d}  \hat{ \chi}^d \right)   \nonumber \\[0.1cm]
 &   &\;=\;&  \text{const}  +    m_D^2 F_D^2 \big( \frac{1}{2} \hat{D}^2 -  \frac{d+\De}{3!} \hat{D}^3  
+  \frac{(d+\De)^2}{4!} \hat{D}^4 +
\ORD(D^5)\big) \;,
\end{alignat}
\comb{with $\hat{D} \equiv D/F_D$}, and a well-defined minimum ($V'(1)=0$ and $V''(1)= m_D^2$). 
The extra term is usually associated with the presence of an operator of scaling dimension $\De$ which breaks 
conformal invariance.  
In \SEC\ref{sec:soft} we will see that in our framework the soft theorem  indicate that $\De =d-2$.  
Indeed the quark-mass term in $\Lagmq|_{\pi=0}$ \eqref {eq:Lagmq} corresponds to the $\De = \Deqq = d-2$ term for which  
 the  Zumino-term \eqref{eq:Zumino}  assumes the form  
  \begin{equation}
 \label{eq:Zumino}
 V_Z(\XXh) =  \frac{\Deqq}{d} \sum_{q=1}^{\Nf}  \mq \vev{qq} \XXh^d \;,
 \end{equation}
 where   $\la \propto \mq$ is a special case of a  symmetry-breaking parameter discussed above.
It assures positivity of  the dilaton mass
 \begin{equation}
 \label{eq:D-GMOR}
  F_D^2 m_D^2 = (d-\Deqq)\Deqq  \frac{\Nf}{2} F_\pi^2 m_\pi^2 \Big|_{d=4,\Deqq=2} = 2 \Nf F_\pi^2 m_\pi^2 \;,
 \end{equation}
 as it contributes the $d\,\Deqq  $-term.
 In analogy to the pion case we will refer to \eqref{eq:D-GMOR} as the \emph{dilaton GMOR-relation}.  It will be derived from a double-soft theorem in  \SEC\ref{sec:soft}. 
 To this end let us summarise the LO 
 \dchiPT Lagrangian considered for $d=4$ ($\Deqq = 2$)
 \begin{alignat}{2}
\label{eq:LO}
& \LagLOdchiPT  &\;=\;&    \frac{ F_\pi^2 }{4}  \hat{\XX}^{2}  \Tr [ \partial^\mu  U \partial_\mu  U^{\dagger}]   +   \frac{1}{2} (\partial\XX)^2 +   \frac{1}{12} \XX^2 R  \quad  + 
 \nonumber \\[0.1cm]
& &\;\phantom{=}\;& \frac{ B_0 F_\pi^2}{2} \big(  \Tr[{\cal M}U^\dagger +U {\cal M}^\dagger ] \,\hat{\XX}^{2} -  \Tr[{\cal M} + {\cal M}^\dagger ]\,  \hat{\XX}^{4} \big)   \;,
\end{alignat}
where the anomalous part has been dropped as per above.   
We refrain from parameterising further potential terms at LO, 
as they are not  required for our work and neither is it clear what form they would take. 
If we want to derive the insertion of the $\bar qq$-operator then it is important 
to realise that its source $s(x)$ is only to be substituted in the term containing $U$ where it acts as a true spurion,  
 ${\cal M}U^\dagger \to ({\cal M}+s) U^\dagger$, and not in the Zumino-term.
We further note that  similar  Lagrangians have frequently appeared in the literature  
\cite{Leung:1989hw,Matsuzaki:2012xx,Crewther:2015dpa,Golterman:2016lsd,Appelquist:2022mjb}. 
The novelty in our case is  the justification for the absence of the anomaly 
term, the   $\Deqq = 3- \gast = 2$ requirement and with respect to many references the inclusion 
of the $ R \chi^2$-term.

As  we sometimes  consider the \chiPT Lagrangian on its own, we quote its LO version
 \begin{equation}
 \label{eq:LOchiPT}
 \LagLOchiPT  =    \frac{ F_\pi^2 }{4}  \Tr [ \partial^\mu  U \partial_\mu  U^{\dagger}]   + 
 \frac{ B_0 F_\pi^2}{2}  \Tr[{\cal M}U^\dagger+U {\cal M}^\dagger ]  \;,
 \end{equation}
  which follows  from \eqref{eq:LO} by setting $\XX \to 1$ and dropping constant terms.

\section{Scalar Correlators  with Goldstones in the Deep-IR ($\mq =0$)}
\label{sec:deep}

In CFTs, correlation functions are determined by a minimal amount of information. 
For example,  two-point functions   are characterised by a single parameter
\begin{equation}
\label{eq:2CFT}
\vev{\Op(x) \Op^\dagger(0) }_\CFT \propto \frac{1}{(x^2)^{\De_\Op}} \;,
\end{equation}
 the scaling dimension $\De_{\Op}$  e.g. \eqref{eq:De} ($\vev{\dots}$ stands for the 
 vacuum expectation value (VEV) hereafter).   For theories 
which flow into an IR fixed point (i.e. conformal window cf. \FIG\ref{fig:CW}), \EQ 
 \eqref{eq:2CFT} represents the leading behaviour in the deep-IR, $x^2 \to \infty$.  
 The aim of this section is to investigate how this  picture is affected in the presence 
 of Goldstone bosons (due to scale and chiral symmetry breaking).  
 The basic reasoning is that since the EFT and QCD describe the same IR physics, the following must hold 
\begin{equation}
\label{eq:match}
 \vev{\Op(x) \Op^\dagger (0) }_{\text{CDQCD}} =   \vev{\Op(x) \Op^\dagger(0) }_{\text{d}\chi\text{PT}} \;, \quad \text{for } x^2 \to \infty \;,
\end{equation}
for the correlation functions as they represent physical observables. 
We will first analyse the two-point functions from the RG viewpoint combining it with some knowledge from 
the EFT described in \SEC\ref{sec:EFT}. 
This will provide a deeper understanding of the analysis in our previous work   \cite{Zwicky:2023bzk} 
and the reason why  the scalar adjoint correlation function was singled out.

Let us introduce operators  $\Op$,  split into ones 
 \begin{equation}
\label{eq:OpsGB} 
 \OS = \bar qq  \;, \quad  \OPa = \bar q     i \ga_5 \, T^aq   \;,
\end{equation}
which couple to single Goldstones and those 
\begin{equation}
\label{eq:Ops} 
 \OSa = \bar q   T^a q\;,  
  \quad  \TT = \frac{\be}{2 g} G^2  \;,
 \end{equation} 
that do not. 
Their quantum numbers  and scaling dimensions are
\begin{equation}
J^{PC}(\OS) = 0^{++}\;, \quad  J^{P}(\OPa) = 0^- \;, \quad J^{P}(\OSa) = 0^+ \;, \quad  J^{PC}(\TT) = 0^{++} \;,
\end{equation}
   and  
\begin{equation}
\label{eq:Dels}
  \De_{\OPa}  =  \De_{ \OS}  =  \De_{ \OSa}  = 3 - \gast   \;,  \quad \De_{\TT} = 4 + \bestp \;,
\end{equation}
respectively. 
The reason of why \eqref{eq:Dels} holds true for the currents can be found in  \cite{Zwicky:2023bzk},  and 
for $\TT$ it is given in \APP\ref{app:conv}.  It is noted that the scaling dimension 
of $G^2$ and  $\TEMT$  are the same, $ \De_{\TT} = \De_{G^2}$, since multiplying 
by the  $\be$-function which is a scalar does not alter the long-distance behaviour of the operator.  
 Two-point functions  will sometimes be abbreviated as 
\begin{equation}
\label{eq:2pt}
\GaE{\Op}(x^2) \equiv \vev{\Op(x) \Op(0) } \;,
\end{equation}
and refer to Euclidean space unless otherwise stated.


\subsection{Renormalisation group  analysis of correlators}
\label{sec:RG}


Since Goldstones are massless, it is to be expected that they will affect the  IR-behaviour.
From the formal RG viewpoint the main change is the presence of an additional scale \cite{DelDebbio:2021xwu}.
 There are in fact two, the pion and the dilaton decay constant but as 
they are of the same order we may group them into one single quantity 
$F = F_{D,\pi}$.  Under these circumstances the RG equation for the correlators 
of the type \eqref{eq:2pt} assume the following  form, e.g. \cite{Cardy:1996xt},
 \begin{equation}
\label{eq:RGE}
(2 \partial_{\ln x^2}  - y_{F^2} \partial_{\ln F^2} + \De_{\Op})\GaE{\Op}(x^2,F^2)  = 0 \;,
\end{equation}
where we have neglected the  $\be \, \partial_{\ln g}$-term since its effect is subleading as will become clear later on. 
Note that  $y_{F^2} = d_{F^2} + \ga_{F^2} = 2$ since $F$ has vanishing anomalous dimension.  
The solution of this equation reads 
\begin{equation}
\label{eq:RGEsol}
\GaE{ \Op}(x^2,F^2)  \propto \frac{1}{(x^2)^{\De_ \Op}} h_{\Op}( x^2 F^2) \;,
\end{equation} 
where in general $h_{\Op}$ is arbitrary such that  predictiveness  is essentially lost.
However, we can improve this situation by matching to \dxchiPT, as in \eqref{eq:match}, taking 
 advantage of the explicit  Lagrangian \eqref{eq:LO}. 
We make the following observations:
\begin{itemize}
\item [a)] We will argue that the EFT cannot produce any  non-integer powers of $1/x^2$. Whereas  
 the EFT  expansion is in powers of $1/(x^2F^2)$  it could be that  
 $\ln x^2$-corrections, related to the neglect of the $\be$-function, 
  resum to non-integer  powers ($n$ integer and $\eta$ not)
\begin{equation}
\label{eq:balance}
\frac{1}{(x^2)^{(n + \eta)}} = \frac{1}{x^{2n}} e^{ -\eta \ln x^2}  =  \frac{1}{x^{2n}} ( 1  - \eta \ln x^2 + \frac{\eta^2}{2} \ln x^2  + \dots ) \;.
\end{equation}
 The answer is however negative since $\eta$ itself must be proportional to inverse powers of $F$, 
  but then there is no other scale in the chiral limit to make 
$\eta$ dimensionless.  
Hence we conclude that $\ln x^2$-terms can only appear in subleading terms which is  natural 
from an EFT perspective  (cf. \SEC\ref{sec:TT} for a more explicit discussion of this aspect). 
Thus  we may write  
\begin{equation}
\label{eq:h}
h_{\Op}(z) =  a_{k_{{ \Op}}} z^{k_{{ \Op}}} + \dots + a_0 + a_{-1} \frac{1}{z} +  \ORD(z^{-2})  + \ln\text{-terms} \;,
\end{equation}
where $k_{\Op}$ is some positive integer and subleading terms contain $\ln x^2$-corrections which 
we have not indicated explicitly.  
\item [b)] Having learned that the overall powers are integers,  
there is still an ambiguity left and that is the  interpretation of the $\De_{\Op}$ coefficient (or the actual number $k_\Op$). 
If the operator $\Op$ shares the quantum numbers with the Goldstone boson $\pi$
and $\matel{\pi}{\Op}{0} \neq 0$, then 
$\GaE{\Op}(x^2)$ scales as 
\begin{equation}
\GaE{\Op}(x^2) \propto \frac{1}{x^2} + \ORD(x^{-4})\;,
\end{equation}
as a consequence of the spectral representation. Are we to interpret $\De_{\Op}$ with this contribution, that 
is $\De_{\Op}=1$? The analysis in terms of the spectral function, further below,  suggests otherwise 
since these contributions are discontinuous with respect to the conformal window phase.  
The RG analysis in \SEC\ref{sec:RGder} provides a further tool. Since the single [two] Goldstone cases 
are of $\ORD(F^2)$ $[\ORD(B_0)]$ with  $\De_{F^2} =2$ and $\De_{B_0} =0$ it is the  latter which agrees with 
the pure CFT scaling.  And thus one is to discard the single Goldstone case. 
Another way to look at it is to the pure CFT case for which the operator product expansion (OPE) provide a strong 
tool. These corrections can then be seen as emerging due to VEVs which is a more formal way to see 
the discontinuity with the conformal window.
\item [c)] There is  a second exception. It could be that in the leading term $\GaE{\Op}(x^2) = c/(x^2)^{\De_\Op}$, 
$c=0$ due to some symmetry.  The identification of $\De_{\Op}$ would then proceed through a next-leading correction 
which complicates matters as is the case for the $\TEMT$-correlator for \dchiPT\!\!. 
\end{itemize}
Finally,  we conclude that  it is the leading multiparticle state contribution which 
is to be identified with $\De_{\Op}$ (provided the exceptional case  under  the last item  does not occur).  

\subsection{Spectral analysis and matching to (dilaton-)\chiPT}
\label{sec:spectral}

Here we aim to support the previous discussion from the viewpoint   of the spectral 
representation which provides  more intuitive insight. 
Below we give a brief summary (e.g. \cite{Weinberg:1995mt,Zwicky:2016lka} for further reading).
In Minkowski space the spectral density $\rho(s)$\footnote{The representation \eqref{eq:GaF} will in general need subtraction terms which 
are not important since polynomial  $(p^2)^n$-terms  corresponds to  local  $\de^{(n)}(x)$-terms which are irrelevant for the 
deep-IR.} 
\begin{equation}
\label{eq:GaF}
\GaF{ \Op}(p^2) \equiv
 i \int d^4x  \, e^{i px}  \vev{T \Op(x) \Op^\dagger(0) }  = 
 \int_0^\infty ds  \, \frac{ \rho_{ \Op}(s) }{s-p^2-i0}  \;,
\end{equation}
of the two-point function is proportional to the imaginary part
\begin{equation}
\rho_ {\Op}(s) = \frac{1}{\pi} \Ima \GaF{ \Op}(s) \;. 
\end{equation}  
The same spectral function enters the Euclidean correlation function 
\begin{equation}
 \int d^4x  \, e^{i px} \vev{ \Op(x) \Op^\dagger(0) } = 
 \int_0^\infty ds  \, \frac{ \rho_{ \Op}(s) } { s+p^2}        \;. 
\end{equation}
It is instructive to  first consider an example of a QCD and a CFT spectral function since the CFT-Goldstone case  
bears aspects of both of them.

\subsubsection*{Spectral function in QCD with a heavy $b$-quark}

In QCD a typical spectral function consists of a few (stable) bound states, 
characterised by $\de$-functions, 
and a continuum, beginning at some branch point  $s_0$, 
\begin{equation}
\label{eq:rhoP}
\rho^{\text{QCD}}_ {\Op}(s) =   \sum_n  |f_{\Op}^{(n)}|^2 \de(s- m_n^2) + \theta(s-s_0) \sig_{\Op}(s) \;,  
\end{equation}
including resonances as well as other multiparticle states. The $\de$-function prefactor is 
$ f_{\Op}^{(n)} \equiv \matel{0}{\Op}{n}$, sometimes referred to 
as the decay constants since such a quantity governs the pion decay.   
We consider  pure QCD, QED and weak interactions turned off,  
with a heavy $b$-quark flavour $\Op(x) \to \bar b i \ga_5 q(x)$
 for which: $m_1^2 \to m_{B}^2$  with no further stable states 
and the continuum threshold is given by $s_0 = (m_{B}+2 m_\pi)^2$. 
The $x$-dependence due to $\sig_{\Op}$ cannot be evaluated without knowing the function but the 
$\de$-term part is simply given by the Fourier transform of the propagator
\begin{equation}
\label{eq:1pt}
\rho_ {\Op}(s)  \propto  \de(s-m^2)   
\quad \Leftrightarrow  \quad\vev{ \Op(x) \Op^\dagger(0) }  \propto
\int   \frac{  d^4 x \,e^{i px} }{p^2 + m^2}\big|_{m^2=0} \propto \frac{1}{x^2} \;,
\end{equation}
which exhibits the   $1/x^2$-scaling  stated earlier.
 
\subsubsection*{Spectral function in a CFT without spontaneous symmetry breaking }

It is straightforward to deduce  that the following identification holds 
\begin{equation}
\label{eq:rhoCFT}
\rho^{\text{CFT}}_ {\Op}(s) \propto s^{\De_{ \Op} - \frac{d}{2}} \quad \Leftrightarrow  \quad 
\vev{\Op(x) \Op^\dagger(0) } \propto  \frac{1}{ (x^2)^{\De_{ \Op}}} \;,
\end{equation}
either by direct computation   \cite{Gelfand:105396} or on grounds of dimensional analysis. 
This function is to be interpreted as belonging to the multiparticle threshold $\sig_{\Op}$.  
It is noted that the limit $\De_\Op \to 1$ for $d=4$  is pathological (IR-divergent spectral integral)  
since the operator really describes a free field rather than a multiparticle spectrum.

\subsubsection*{Spectral function in spontaneous scale symmetry breaking}

The case of  spontaneous  symmetry breaking has elements of both the QCD- and the CFT-case.  
The stable particle becomes the massless Goldstones and the continuum threshold 
moves to zero  assuming a simple power law. 
The $\de$-functions are very singular and have no counterpart in the unbroken CFT-case and it is 
intuitively clear that they should not be identified with $\De_{\Op}$. 
In \SEC\ref{sec:RG} this has been argued more formally, that $\De_{\Op}=1$ 
does strictly imply a free field and is therefore not associated with $s$ raised to some power. 
In conclusion the spectral function, in the case of  SSB, generically reads
\begin{equation}
\label{eq:rhoSSB}
\rho^{\text{SSB}}_ {\Op} \propto s^{{\De}_{\Op}-\frac{d}{2}} \left( F^2 \de(s)  
+ c\,  \theta(s) \right) + \de \rho(s)  \;,
\end{equation}
where $c$ is a constant  and the $\de$-term is only present if the Goldstone couples to $\Op$.  
The quantity $\de \rho(s)$ parameterises   suppressed contributions such as multi-nucleon thresholds.

\subsubsection{Multi-Goldstone case:  the  operator $\OSa$}
\label{sec:no1GB}

The case of the operator $\OSa$ \eqref{eq:Ops}  is the simplest as its quantum numbers 
 do not allow for the propagation of a single Goldstone.  
 It was chosen for this reason in our earlier work   \cite{Zwicky:2023bzk}
which  we recapitulate this in the language of the spectral function.  
Dropping the $\de$-function piece in \eqref{eq:rhoSSB} one gets the standard CFT-scaling
\begin{equation}
\rho_ {\OSa}(s) \propto s^{\De_{ \OSa} - \frac{d}{2}} \quad \Leftrightarrow  \quad 
\CoE{ \OSa}_{\CDQCD} \propto  \frac{1}{ (x^2 )^{\De_{ \OSa}}} \;.
\end{equation}  
In  \dchiPT these operators are matched at leading order to  
\begin{equation}
\label{eq:JSa}
S^a|_{\LO} = - \frac{F_\pi^2 B_0}{2} \Tr[T^a U^\dagger + U T^a]\hat{\XX}^2  = \frac{B_0}{2} d^{abc} \pi^b \pi^c +  \ORD(1/F) \;,
\end{equation}
to two pions using the method of sources  \cite{Zwicky:2023bzk}.
Its evaluation involves the propagation of two free pions $\vev{\pi^a(x) \pi^b(0)} = \frac{\de^{ab}}{4 \pi^2 x^2}$ 
only, as illustrated in \FIG\ref{fig:dias}, and one gets 
\begin{equation}
\CoE{ \OSa}_{\chi \text{PT}} \; \propto \; \frac{1}{x^4}   \;.
\end{equation}
 Matching to \eqref{eq:match}, $\CoE{ \OSa}_{\text{CDQCD}}  \propto \frac{1}{(x^2)^{\De_{\OSa}}}$ one finds 
\begin{equation}
 \De_{\OSa} = 3- \gast  =2 \quad \Leftrightarrow \quad \gast =1 \;,
\end{equation}
that the mass anomalous dimension at the IR fixed point is unity  \cite{Zwicky:2023bzk}.

\subsubsection{Single-Goldstone case:  the operators  $\OPa$ and  $\OS$ }
\label{sec:1GB}

\begin{figure}[h!]
\includegraphics[width=0.9\linewidth]{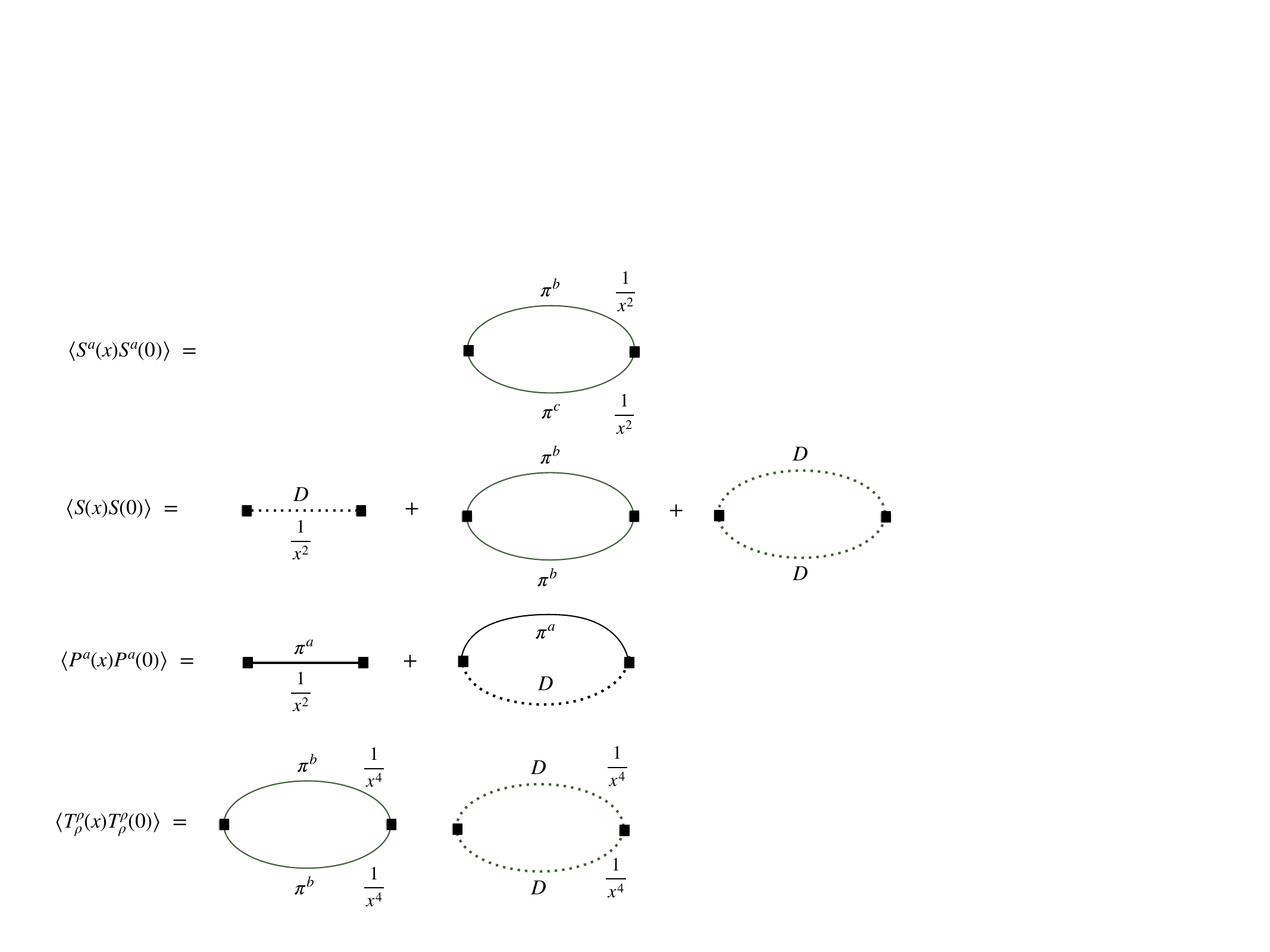}
\vspace{-0.1cm}
	\caption{\small  
	Two-point  functions of operators defined in (\ref{eq:OpsGB},\ref{eq:Ops}) evaluated in the deep-IR. 
	 (top) scalar adjoint correlator with no coupling to Goldstones scaling as $1/x^4$. (centre) 
	 scalar and adjoint pseudoscalar correlators with a single and a two  dilaton intermediate state. 
	 (bottom) $\TEMT$-correlator in \chiPT (the one for \dchiPT is zero at LO).}
	\label{fig:dias}
\end{figure}

The cases of the operators  $\OPa$ and  $\OS$  \eqref{eq:OpsGB} differ  in that they 
couple to a single Goldstone, the pion and the dilaton respectively, and thus both terms in 
\eqref{eq:rhoSSB} are present. Omitting the truly subleading terms we have
\begin{equation}
\rho_ {\OPa}(s) \propto  
s^{{\De}_{\OPa}-\frac{d}{2}} ( F_\pi^2 \de(s)  +  c  )    \;,
\end{equation}
implying 
\begin{equation}
\CoE{ \OPa}_{\CDQCD} \propto \frac{F_\pi^2}{  (x^2 )^{\De_{ \OPa}-1}} +   \frac{c'}{ (x^2 )^{\De_{ \OPa}}} \;,
\end{equation}
where $c$ and $c'$  are constants. 
The case of $\OS$ is analogous with $F_\pi \to F_D$, cf.\,\FIG\ref{fig:dias}. 
In \dchiPT the operators are given by
\begin{alignat}{3}
\label{eq:spec}
 &  S   &\;=\;&  -  \frac{F_\pi^2 B_0}{2}  \Tr[ U^\dagger + U ] \hat{\XX}^2  &\;=\;& \phantom{-} 2 B_0 \Nf F_\pi^2  \hat{D}  +   \frac{B_0}{4} (  \pi^a \pi^a - \Nf {F}_\pi^2  \hat{D}^2 )  +  \ORD(1/F) \;,    \nonumber \\[0.1cm]  
 &  P^a   &\;=\;&  - \frac{F_\pi^2 B_0}{2} i \Tr[T^a U^\dagger - U T^a] \hat{\XX}^2  &\;=\;& 
 - B_0 F_\pi \pi^a (1 - 2 \hat{D})
  +  \ORD(1/F) \;.
\end{alignat} 
 From these expressions the following LO correlation functions result 
\begin{alignat}{2}
& \CoE{ \OS}_{\text{d}\chi\text{PT}}  &\;=\;&   \frac{c_S^{(2)}}{  x^2} + 
 \frac{c_S^{(4)}}{x^4}      \;,\nonumber \\[0.1cm]  
& \CoE{ \OPa}_{\text{d}\chi\text{PT}}  &\;=\;&  \frac{c_{P^a}^{(2)}}{x^2}
 +   \frac{c_{P^a}^{(4)}}{x^4}     \;. 
\end{alignat}
with
\begin{equation*}
c_S^{(2)} =   \frac{ B_0^2 F_\pi^2 \hat{F}_\pi^2  \Nf^2}{\pi^2} \;, \quad
c_S^{(4)} =     \frac{B_0^2  ( (\Nf^2-1)+ \Nf^2 \hat{F}_\pi^4) }{32 \pi^4 } \;, \quad 
c_{P^a}^{(2)} =  \frac{ B_0^2 F_\pi^2}{4 \pi^2}  \;, \quad
c_{P^a}^{(4)} =  \frac{B_0^2 \hat{F}_\pi^2}{4 \pi^4} \;.
\end{equation*}
Here and thereafter the flavour index is understood to be held fixed. 
Matching the two expressions, as in \eqref{eq:match} but disregarding the $1/x^2$-contribution as argued above, one deduces 
\begin{equation}
 \De_{\OS} =  \De_{\OPa} = 3- \gast  = 2 \quad \Leftrightarrow \quad \gast =1 \;,
\end{equation}
the same result as in the previous section.   This should not be taken for granted but as a sign of the consistency 
of the approach.

\subsubsection{Multi-Goldstone case II: the trace of the energy momentum tensor  $\TEMT$}
\label{sec:TT}

The case of the TEMT is particularly interesting  as it brings several new elements.  
It is tempting to think that  the dilaton would couple 
to $\TEMT = \frac{\be}{2g} G^2$ but it does in a way \eqref{eq:FDTmunu}
\begin{equation}
\matel{D}{\TEMT}{0} = F_D m_D^2 \;,
\end{equation}
which vanishes when the dilaton is massless, if it is massive,  it   decouples from the IR physics.  
The next candidates in question are  two pions and two dilatons.   
 It is instructive to first consider the gluon correlator for which $\De_{G^2} = 4 + \bestp$ \eqref{eq:De}   
 is determined by the  slope of the $\be$-function at the fixed point
 \begin{equation}
\rho_ {G^2}(s) \propto s^{\De_{ G^2} - \frac{d}{2}} \quad \Leftrightarrow  \quad 
\CoE{ G^2}_{\CDQCD} \propto   \frac{1}{ (x^2 )^{4+ \bestp}}   \;.
\end{equation}
Adapting it to $\TEMT$ one  multiplies two powers of the  $\be$-function which as argued previously,  leaves the 
$x$-scaling unaltered 
\begin{equation}
\label{eq:TTbe}
\CoE{\TEMT}_{\CDQCD} \propto   \frac{\be^2}{ (x^2 )^{4+ \bestp}}  \;.
\end{equation}
In the EFT it makes a difference  whether one uses   \chiPT  \eqref{eq:LOchiPT} or   \dchiPT in \eqref{eq:LO}.  From the LO EFTs one obtains \cite{Zwicky:2023fay}, neglecting terms of higher powers in $\pi$,
\begin{equation}
\label{eq:improved}
\TEMT|^\LO_{\chi\text{PT}}  = - \frac{1}{2} \partial^2 \pi^a \pi^a  \;, \quad 
\TEMT|^\LO_{d\chi\text{PT}}  = 0   \;.
\end{equation}
Thus at LO \chiPT is only scale invariant but not conformal contrary to \dchiPT\!\!.
One gets 
\begin{equation}
\label{eq:TTLO}
\CoE{ \TEMT}^\LO_{\chi\text{PT}}  \;\propto\; \frac{1}{x^8}  \;, \quad 
\CoE{ \TEMT}^\LO_{\text{d}\chi\text{PT}}  \;\propto\; 0   \;.
\end{equation}
In \chiPT the matching with  \eqref{eq:TTbe} implies,  in straightforward manner,  that 
the slope of the $\be$-function vanishes at the IR fixed point: $\bestp=0 $.
The case of \dchiPT is exceptional and corresponds to the case discussed under item c) in \SEC\ref{sec:RG} 
in that the LO-term vanishes by conformal symmetry.   
In fact one can show that $\vev{G^2(x)G^2(0)}$ has no $1/x^8$-term as otherwise the $\be$-function ought to 
vanish entirely. Thus one cannot argue  $\bestp=0$ by expanding in $\de g$.  
We proceed in a different manner  using RG arguments. 

\subsubsection{Renormalisation group derivation of $\bestp=0$}
\label{sec:RGder}

We first consider the GMOR-relation and the TEMT to deduce $\matel{\pi}{G^2}{\pi} = \ORD(\mq)$ and by equating 
to the RG version of the matrix element we deduce $\bestp=0$.

We may decompose the TEMT as  $\TEMT =  \Oone' + 2 \Otwo$  
 with $\Oone' \equiv \Oone - \Otwo $   \eqref{eq:inv1}.  Note that  $\Oone'$ inherits the  RG invariance of $\Op_{1,2}$. 
  Since $2 \Otwo$ saturates the GMOR-relation $\matel{\pi}{\TEMT}{\pi} = \matel{\pi}{2 \Otwo}{\pi} + \ORD(\mq^2)$,  
 one gets that the $\Oone'$ matrix element must vanish to linear order in $\mq$. Expanding  
  in $\de g$, assuming $\gast =1$ and $\best=0$, we get 
\begin{equation}
\label{eq:O1p}
0 = \matel{\pi}{\Oone'}{\pi}|_{\mq} =  \de g \Big(  \frac{\bestp}{2 g_*}   \matel{\pi}{G^2}{\pi}   + \sum_q \gastp \mq  \matel{{\pi}}{\bar qq}{{\pi}} \Big)|_{\mq} + \ORD((\de g)^2 ) \;.
\end{equation}
Without making any assumptions on $\bestp$ and $\gastp$ we deduce that $\matel{\pi}{G^2}{\pi} = \ORD(\mq)$ 
since we know that  $\matel{{\pi}}{\bar qq}{{\pi}} = \ORD(\mq^0)$ from the GMOR-relation.  
From general RG arguments, similar to hyperscaling, e.g.   
\cite{Cardy:1996xt,DelDebbio:2010ze,DelDebbio:2010jy,DelDebbio:2013qta} and also \cite{DelDebbio:2021xwu}, 
one infers the $\mq$ and the  $F$ dependence
\begin{equation}
\label{eq:G2RGa}
{\matel{\pi}{G^2}{\pi} =  \mq^{\frac{\De_{G^2} + 2 d_\pi}{y_m}} h( {F} {m}_q^{-1/y_m}   ,\mu) } \;,
\end{equation}
where $\mu$ is the RG scale, $y_m \equiv 1 + \gast$, {$d_\pi = -1$ is the dimension of $\state{\pi}$  \cite{DelDebbio:2010jy}}
and $h$  an  a priori unknown function.  
In principle there is also the parameter  $B_0 = - \vev{\bar qq}/F_\pi^2$ but it can be omitted since its
scaling dimension vanishes   $\Delta_{B_0} = \Delta_{\bar qq} - \Delta_{F_\pi^2} =0$. 
Thus the extra powers of $F^2$ is what distinguishes the scaling in the QCD IR-phase  from the conformal 
window phase.  

Hence, our task is to find this power in $F$. 
We can do so by using that $\matel{\pi}{G^2}{\pi}$ and $\matel{\pi}{\bar qq}{\pi}$ are of the same order in $F_\pi$. 
From the soft-pion theorem one deduces  $\matel{\pi}{\bar qq}{\pi} \propto \frac{1}{F_\pi^2}  \vev{\bar qq}  = -B_0 $.  
{ Thus we conclude that $\matel{\pi}{G^2}{\pi} \propto F^0$ and we obtain 
\begin{equation}
\matel{\pi}{G^2}{\pi} \propto   \mq^{\frac{2+ \bestp }{y_m}}   \;,
\end{equation}
}and equating to $\matel{\pi}{G^2}{\pi} = \ORD(\mq)$  deduced above,
one infers  consistency (using $y_m=2$) if and only if the slope vanishes 
\begin{equation}
\label{eq:bestp}
\bestp=0 \;.
\end{equation}
Reassuringly, this matches the result obtained  previously for \chiPT\!\!.
 
 To gain 
further confidence consider $\vev{\bar qq}  = -B_0 F_\pi^2$ in terms of an RG analysis as in \eqref{eq:G2RGa}. 
Interpreting  $\vev{\bar qq} \propto F^2$, thereby ignoring  $B_0$, we may write 
\begin{equation}
 \vev{\bar qq}   = \mq^{\frac{\Deqq}{y_m}} H({F} {m}_q^{-1/y_m}  ,\mu) \propto  F^2 \mq^{\frac{\Deqq-2}{y_m}}   \propto  \ORD(1) \;,
\end{equation}
which yields the correct result:  no quark-mass dependence at LO with $\Deqq =2$.

 \subsection{Consequences of $\bestp=0$}

In our view, $\bestp=0$  is an important result. 
\begin{itemize}
\item [a)] It is   supported by smooth matching to $\NNone$ supersymmetry (cf. \SEC\ref{sec:SUSY}).
\item [b)] It is the  promised justification for dropping  the anomaly term  
 in the LO  Lagrangian \eqref{eq:LO0}.  For example, in \REF \cite{Cata:2018wzl}
the LO anomaly terms become 
degenerate with the kinetic terms in the limit $\bestp \to 0$, and may therefore be dropped. 
 As   \chiPT is an EFT the EMT needs renormalisation \cite{Donoghue:1991qv}. In that case 
 three new  counterterms,  parameterised as a function of the Riemann tensor, have 
 been found  at NLO.  
 We expect  the same program  
 to apply in \dchiPT such that the breaking of scale invariance emerges naturally at higher orders.\footnote{In \dchiPT  the 
divergent parts of the counterterms have been computed very recently \cite{Freeman:2023ket} 
but not  the ones for the EMT.} We would though expect that the terms from the Weyl or conformal anomaly \cite{Schwimmer:2010za}, which are the analogue of the Wess-Zumino-Witten term in QCD, would need to be matched 
at NLO. 
\item [c)] The slow logarithmic running (discussed below) 
may be interpreted in terms the mass gap amongst the hadrons.  
Once the $\rho$, $a_0$-meson, depending on the channel,  decouple
the theory asymptotes slowly into the free Goldstone EFT (accompanied by $1/(x^2F^2)$ power  correction at NLO in accordance with 
the earlier discussion). 
\item [d)] It seems to make the existence of a QCD Seiberg dual in the non-supersymmetric conformal window \cite{Terning:1997xy,Sannino:2009qc,Mojaza:2011rw} more likely since the turnaround of the $\bestp$ in the conformal window 
would go well with a weakly coupled dual IR fixed point. 
\end{itemize}
  Moreover, as emphasised in \cite{Rho:2023vow,Shao:2024qny}, $\bestp=0$ also impacts on the nuclear axial current 
  which has been the subject of scrutiny for many decades.  In the light of  $\bestp=0$ it might be of interest to 
  determine the corrections due to the strange quark mass.

\subsubsection{Logarithmic running}

We may expand the $\be$-function around  the fixed point coupling $\de g \equiv g - g_*$
\begin{equation}
\label{eq:beIR}
\be = \bestp \de g+ \frac{1}{2} \bestpp {(\de g)^2} + \ORD( (\de g)^3) \;.
\end{equation}
By the very assumption of an IR fixed point we have $\bestp \geq 0$ and  the value in \eqref{eq:bestp} saturates this constraint. Using  $\bestp =0$ gives an RG equation 
\begin{equation}
\frac{d}{d \ln \mu} \de g = \frac{1}{2} \be_*^{''} \de g^2 \;,
\end{equation}
which  solves to a logarithmic  
\begin{equation}
\label{eq:log}
\de g(\mu) 
 = \frac{4}{ |\be_*^{''} |\ln \frac{\mu^2}{\LaQCDIR^2}}  \;,
\end{equation}
instead of to a power-like form $\de g \propto \mu^{\bestp}$. 
Above, the boundary condition $\de g(0) = 0$ was imposed and 
the  expression is maximally valid  for $\mu < \LaQCDIR$  where it runs into a Landau pole from below. 
The scale  $\LaQCDIR$ is the analogue of $\LaQCD$ in the IR.
Note that whereas 
  $\bestp$ is scheme independent under regular coupling 
redefinitions this is not the case for $\bestpp$.  For the sake of concreteness we may assume it to be non-vanishing, 
in which case it needs to be negative to assure negativity of the $\be$-function \eqref{eq:beIR}.
In retrospect the logarithmic  running is reassuring as it is difficult to see how 
\dxchiPT  could reproduce power-like scaling as per item c) above.


\subsubsection{The vanishing of $\bestp$ implies the vanishing of $\gastp$}
\label{sec:same}

In ${\cal N}=1$ supersymemtric gauge theories  $\bestp$  is proportional to $\gastp=0$ 
 due to the NSVZ $\be$-function \cite{Shifman:2023jqn}.   
 Hence, the vanishing of $\bestp$ implies the vanishing of $\gastp$ and it is natural to ask whether this extends
 to the non-supersymmetric case.  Using \EQ\eqref{eq:O1p} one infers that
zero $\bestp$ implies zero $\gastp$ by RG invariance:
\begin{equation}
\label{eq:gastp}
\bestp = 0 \quad \Leftrightarrow \quad  \gastp=0 \;.
\end{equation}
We note that this result might be more universal since they both 
govern the correction to hyperscaling e.g. \cite{DelDebbio:2013qta}.

\section{Comparison with ${\cal N}=1$ Supersymmetric  Gauge Theories}
\label{sec:SUSY}

It is instructive to reflect on the results obtained in the context of  ${\cal N}=1$ supersymmetric gauge
theories and its Seiberg dualities, for which many excellent reviews are available  
\cite{Intriligator:1995au,Shifman:1999mk,Terning:2006bq,Tachikawa:2018sae}.   
Our brief summary  below,  extends  on the non-supersymmetric conformal
window in \SEC\ref{sec:CW}. 
For $SU(N_c)$ gauge theories with $N_f$ fermions, which is referred to as the electric theory, 
the conformal window extends over the following region  
\begin{equation}
\label{eq:freeM}
\text{conformal window:}  \qquad  \frac{3}{2} \Nc \leq   \Nf  \leq 3 \Nc  \;. 
\end{equation}
In this window there exists a Seiberg  dual, referred to as the magnetic theory, which is also asymptotically 
free with gauge group $SU(\Nf - \Nc)$ and a  Yukawa-superpotential interaction 
${\cal W} = \la M_{\bar{j}}^i q_i \tilde{q}^{\,\bar j}$ which couples  
$\Nf$ chiral  $q_i$ and $\Nf$ antichiral  $\tilde q^{\bar{j}}$ matter fields to  
a colour neutral  chiral meson superfield $ M_{\bar{j}}^i$ (where $i$ and $\bar j$ are  $SU(\Nf)_L$ and $SU(\Nf)_R$
flavour indices respectively).   
The duality \cite{Seiberg:1994pq}, in part based on non-trivial matching of anomalies,  is the statement that  both of these theories flow to the same CFT in the IR.  
The relation of the dilatation and non-anomalous $R$-current in the same superconformal multiplet 
allows to determine the scaling dimension of the bilinear (s)quark fields in terms 
of the known $R$-charges\footnote{In the supersymmetry literature $\ga_{Q}$, the 
matter-field anomalous dimension, is used rather than $\ga_m$. They are related by $\ga_m = - \ga_Q$ through 
non-renormalisation theorems to all order in perturbation theory.}
\begin{equation}
\label{eq:rel}
\De_{\tilde Q Q} = 2 - \gast = 3 -  \frac{3\Nc}{\Nf} \;, \quad \De_{\bar qq } = 3 - \gast = 4 -  \frac{3\Nc}{\Nf} \;,
\end{equation}
valid in the conformal window.   
Alternatively, this formula can be deduced by requiring 
the NSVZ $\be$-function \cite{Novikov:1983uc,Novikov:1985rd} to vanish in the domain \eqref{eq:freeM}. 

Below the conformal window squark bilinears can acquire VEVs and break the superconformal 
$R$-symmetry  and the relations \eqref{eq:rel} cease to be valid. 
In the range 
\begin{equation}
\label{eq:IRfreeM}
\text{IR-free magnetic:}   \qquad N_c+1  <  N_f  <  \frac{3}{2} N_c  \;,
\end{equation}
the magnetic theory plays the role of the weakly coupled  IR EFT of the strongly coupled electric theory. 
The magnetic theory is not asymptotically free anymore and the electric theory can be seen as its UV completion. 
We may think of this magnetic description as a meson EFT of the hybrid type, as it still contains coloured degrees
of freedom, and as such resembles the  phenomenologically successful chiral quark model \cite{Manohar:1983md}. 
There are three further phases below: s-confinement for $\Nf = \Nc+1$ (confinement without chiral symmetry breaking), 
$\Nf= \Nc$ anomalies are matched by mesons and baryons only, $\Nc >  \Nf > 1 $ 
has a runaway potential (no stable vacuum). 
We will not focus on these phases as they are quite possibly peculiar to supersymmetry itself. Further note, that for 
 $\Nf = \Nc+1$  the dual gauge group in particular ceases to make sense but interestingly 
 the anomalies can still be matched. 
We discuss $\gast =1$, $\bestp = \gastp=0$  under items a) to c)  below from 
the supersymmetric-viewpoint reasoning whether they extend into the IR-free magnetic phase  \eqref{eq:IRfreeM}. 
\begin{itemize}
\item  [a)]  
The squark bilinear $\tilde  Q Q$ is believed to match onto the free 
meson field $M_{\bar{i}}^j$ for which $\vev{M (x) M(0)} \propto1/x^2$  for $x^2 \to \infty$ must  hold in the IR-free magnetic range \eqref{eq:freeM}. 
 This implies $\De_{M} =  \De_{\tilde  Q Q} = 2 - \gast = 1$ and thus $\gast = 1$ must hold throughout the IR-free magnetic phase.\footnote{As a side remark, 
 it is curious that in ${\cal N}=1$ the end of the conformal window coincides with the unitarity bound whereas for 
  QCD gauge theories  the unitarity bound $\gast \leq 2$ is not reached. 
The interpretation that suggests itself \cite{Zwicky:2023bzk} is  that chiral symmetry breaking sets in once the scaling dimension allow the operator 
$S^a$ ($\De_{S^a} =2$) to produce two free pions as predicted by \dxchiPT\!\!.} 
\item  [b)]  In the conformal window $\bestp$ is non-zero. This can be established close to the  
electric Caswell-Banks-Zaks  fixed point as its very idea is 
based on  $\bestp$ being small. 
As the coupling increases so does $\bestp$, e.g. \cite{Ryttov:2017kmx}.
Since $\bestp$ is also the scaling exponent of the TEMT-correlator  \eqref{eq:TTbe} \cite{Shifman:2023jqn} 
this implies the equality of 
the electric and magnetic slopes in the conformal window\footnote{This result was  derived earlier by matching the Konishi currents via an involved 
 Kutasov construction in  \cite{Anselmi:1996mq}.} 
\begin{equation}
\label{eq:duality}
\best'|_{\el} =  \best'|_{\ma}  \quad \Leftrightarrow  \quad 
  \CoE{\TEMT}_{\el} \quad  \stackrel{\text{IR} }{\longleftrightarrow } \quad  \CoE{\TEMT}_{\ma} \;.  
\end{equation}
The value of $\best'|_{\ma}$ corresponds  to the minimal eigenvalue of the gradient matrix of the 
$\be$-function matrix in the gauge-Yukawa  coupling space. Since the magnetic theory is weakly coupled 
at the lower edge of the conformal window, $\Nf$ towards $\frac{3}{2}\Nc$, where it assumes its own 
perturbative fixed point this means that 
\begin{equation}
\label{eq:bestpCW}
\bestp|_{\Nf=3} = \bestp|_{\Nf=\frac{3}{2}} = 0\;, \quad  \bestp|_{\frac{3}{2} \Nc <   \Nf  < \Nc}  > 0 \;,
\end{equation}
$\bestp$ is zero at both edges of the conformal window and positive in between which   
follows from the very assumption of an IR fixed point. 
This is curious as both, very weak and very strong coupling, seems to makes $\bestp$ vanish. 
However,  since $\bestp =0$ also implies the logarithmic running implicit in the EFT cf. \SEC\ref{sec:TT}, 
this can be seen as a necessary result and matches the expectation of the IR-free magnetic phase \eqref{eq:IRfreeM}. 
This is also the reason it should continue to hold in the IR-free magnetic phase which is though 
not argument inherent to supersymmetry.
\item  [c)] The NSVZ $\be$-function \cite{Novikov:1983uc,Novikov:1985rd} is expressed in terms 
of known quantities and the anomalous dimension $\ga_m$. Hence,  differentiating, and using $\best=0$,  
give a relation    \cite{Shifman:2023jqn}
\begin{equation}
\beta^\prime_* =  \frac{\alpha^2_*}{2\pi}\,\,  \frac{N_f}{1- \frac{\alpha_*}{2\pi}N_c}  \gastp \;,
\end{equation}
 which must hold, at least, in the conformal window.  ($\al_*= g_*^2 /4\pi$ denotes the electric IR fixed point gauge coupling).  Therefore,  $\bestp=0$ at $\Nf = \frac{3}{2} \Nc$ implies the same for 
 $\gastp=0$ at this point.  In IR-free magnetic phase, $\gastp=0$ might continue to hold 
  by the same reasoning as given in \SEC\ref{sec:same}. This argument is not specific to supersymmetry. 
\end{itemize}
In summary, we have argued that for ${\cal N}=1$ supersymmetry  
$\gast =1$ and $\bestp = \gastp=0$ hold at the boundary of the conformal window. 
Using Seiberg duality we provided an argument of why $\gast =1$ and $\bestp=0$ 
carry over into the IR-magnetic phase.
For  $\gastp=0$, its link to $\bestp$ which is not supersymmetric in nature was invoked with 
 regards to the  IR-magnetic phase. 
It is worthwhile to emphasise that there have been interesting attempts to understand the magnetic dual in terms  
of hidden local symmetry and  low energy Goldstone physics   \cite{Komargodski:2010mc,Abel:2012un}. It may well 
be that the massless $0^{++}$ flavour-singlet found in these cases is the dilaton.

\section{  Soft Dilaton  Theorems}
\label{sec:soft}

In this section we apply the double-soft theorem to the matrix element 
\begin{equation}
\label{eq:start}
2 m_D^2 =  \matel{D}{\TEMT}{D} \;, 
\end{equation}
which will provide some surprising model-independent results  and the dilaton GMOR-relation  \eqref{eq:D-GMOR}. 
\EQ\eqref{eq:start} is usually considered a first-principles formula but in 
the presence of  a massless dilaton it does not hold for a standard hadronic state $\varphi$. 
The dilaton pole cancels the $2 m_\varphi^2$-contribution from the kinetic term  such that  
$\matel{\varphi}{\TEMT}{\varphi} =0$ \cite{DelDebbio:2021xwu}. However, since here  $m_D \neq 0$ is kept, 
this situation does not arise,  as we have explicitly checked  at  LO. All the information needed 
from the potential is the mass term $V  \supset \frac{1}{2} m_D^2 D^2$.
 
The idea of soft theorems is that one has a pseudo-Goldstone with a light mass, $m_{\text{pGB}} \ll \LaQCD$, whose momenta can be approximated to be soft $q \to 0$ while keeping $m_{\text{pGB}} \neq 0$ \cite{Donoghue:1992dd} (which  
is automatic in an EFT treatment).\footnote{That the soft theorem is encoded 
in the Lagrangian is of no coincidence  as effective Lagrangians are considered 
a more transparent way to organise them \cite{Weinberg:2009bg}.}
The procedure assumes that  the original matrix element does not change significantly in the soft limit.
The gain is that the evaluation then proceeds by a simpler matrix element where the Goldstone 
emerges in a computable commutator.
For the dilaton and pion they  read 
  \begin{alignat}{3}
\label{eq:soft2}
& \matel{D(q) \be}{{\cal O}(0)}{\al}  &\;=\;& - \frac{1}{F_D} \matel{ \be}{i [Q_D,{\cal O}(0)]}{\al}  &\;+\;&  \lim_{q \to 0} i q \cdot R \;, 
\nonumber \\[0.1cm]
& \matel{\pi^a(q) \be}{{\cal O}(0)}{\al}  &\;=\;& - \frac{i}{F_\pi} \matel{ \be}{[Q_5^a,{\cal O}(0)]}{\al}  &\;+\;& \lim_{q \to 0} i q \cdot R^a \;,  
\end{alignat}
where  the one for pions can be found in  the textbook  \cite{Donoghue:1992dd}.
The $R$'s are the so-called remainders 
   \begin{alignat}{2}
\label{eq:remainder}
&  R_\mu &\;=\;&    - \frac{i}{F_D}\int d^dx e^{i q\cdot x} \matel{ \be}{T J^D_{\mu}(x) {\cal O}(0)}{\al}  
\;,  \nonumber \\[0.1cm]
&  R^a_\mu &\;=\;&    - \frac{i}{F_\pi}\int d^dx e^{i q\cdot x} \matel{ \be}{T J_{5\mu}^a(x) {\cal O}(0)}{\al}  
\;, 
\end{alignat}
which vanish unless there are  intermediate states degenerate with either 
$\al$ or $\be$ \cite{cite-key}.\footnote{We have checked that they vanish in the cases at hand for which 
it is important to keep $m_{D,\pi} \neq 0$.  An example of where the $R$ matters is sketched in the appendix 
of \REF\cite{Zwicky:2023bzk}.}   
Let us focus on an operator present in the TEMT   $\Op \subset \TEMT   $ which is responsible for generating 
the mass.  We will refer to this operator as the mass operator. 
Applying the soft theorem \eqref{eq:start} one gets 
\begin{equation}
\label{eq:partial}
 2 m_D^2 =  \matel{D}{\Op(x)}{D}  = 
 -\frac{1} {F_D} \matel{0}{i [Q_D ,\Op(x)]}{D} =  -\frac{1} {F_D} (  \De_{\Op}  + x \cdot \partial) \matel{0}{\Op (x)}{D}  \;.
\end{equation}
Above  we used the CFT definition of applying $Q_D$ to a primary operator which is an assumption similar 
to \eqref{eq:IRCFT1}.  
The main technical point is as follows.
It is  crucial to keep  the derivative term since  
the matrix element, $\matel{0}{\Op(x)}{D(q)} = F_\Op e^{-i qx}$, carries $x$-dependence. 
Thus the question of how to make sense of this term?  
We may regard this matrix element as a test function in a distribution space and are therefore 
leads to integrate over space by 
\begin{equation}
\mathbb{1}_V = \frac{1}{V} \int_V d^d x \;.
\end{equation}
This allows 
for integration by parts such that\footnote{The proper way to do this would be to form a wave packet  
within a finite region in $x$-space. In addition this makes it clear that the boundary terms that arise 
upon integration by parts do not contribute.}
\begin{equation}
\mathbb{1}_V  [ x \cdot \partial  \matel{0}{\Op (x)}{D} ]  = - d  \,  \frac{1}{V} \int_V d^d x \matel{0}{\Op(x)}{D} \;,
\end{equation}
and assembling we get the \emph{single-soft dilaton theorem}
\begin{equation}
\label{eq:result1}
m_D^2 F_D =  \frac{1}{2}(d- \De_\Op)   \matel{0}{\Op}{D} \;, 
\end{equation}
where the integral has been removed since the second dilaton is to be made soft as well. 
Applying the procedure once more one gets the \emph{double-soft dilaton theorem} 
\begin{equation}
\label{eq:result2}
m_D^2  F_D^2 =  \frac{1}{2} (\De_\Op-d) \De_\Op  \vev{\Op } \;,
\end{equation}
where this time the derivative term can be dropped since $ \vev{\Op(x) }$ has no $x$-dependence by translation invariance. 
It is worthwhile to stress that these two relations are model-independent by which we mean not particular to the 
gauge theory. 

\subsection{Consequences of the single- and the double-soft dilaton theorem}

There are a number of things that one can learn from these two relations.
Only items c) and d) are specific to the gauge theory; a) and b) are  general. 
\begin{itemize}
\item [a)] One has $ \vev{\Op} \leq  0$ necessarily,  
such that $\vev{\TEMT}$ is lowered ($\Op \subset \TEMT$) with respect to the perturbative vacuum. 
Hence, the mass squared \eqref{eq:result2} is indeed 
positive thanks to the $d$-term that originates from the derivative term.  This gives us confidence that this term 
is present 
(cf. item c) for a further comment). 
\item  [b)] One can get a very similar relation to \eqref{eq:result1} by contracting 
\eqref{eq:FDTmunu} 
\begin{equation}
\label{eq:famous}
 m_D^2  F_D   =    \matel{0}{\Op}{D}  \;,
\end{equation} 
where we have assumed $\TEMT \to \Op$ since $\Op$ is the operator that generates the mass.
 Inspecting 
\eqref{eq:famous} and \eqref{eq:result1}
one infers that 
\begin{equation}
\label{eq:DeOp}
\De_\Op = d-2 \;,
\end{equation} 
must hold which seems important. 
Hence,  the soft  theorem indicates that the dilaton can only get a mass from an operator of dimension 
$\De_\Op = 2$ (in $d=4$). 
Alternative derivations of this results are given in \APPs \ref{app:Lag} and \ref{app:massOp} 
directly from the Lagrangian and from scaling arguments. 

On another note, it is tempting to read the results \eqref{eq:DeOp} backwards and interpret 
it as yet another demonstration that $\gast=1$ ($\Deqq = 3-\gast$) has to  necessarily hold. 

\item [c)]  Let us turn to the gauge theory where the operator 
\begin{equation}
\label{eq:Op}
 \TEMT  \supset \Opqq  =  (1+ \gast) \sum_q  \mq \bar qq  \;, \quad 
\end{equation}
satisfies the criteria   \eqref{eq:DeOp} for $\gast =1$ ($\Deqq = 3-\gast$).  It is in fact tempting to read it the
other way around. Namely, as another demonstration that $\gast=1$ must hold. 
Turning to pragmatic matters, one may use \eqref{eq:result2}  to obtain
\begin{equation}
\label{eq:D-GMORa}
 F_D^2 m_D^2 =  \Deqq(\Deqq-d)  \frac{\Nf}{2} (1+ \gast) \mq \vev{\bar qq} = - 4 \Nf  \mq \qq
\;,
\end{equation}
where   $\gast=1$  and $d=4$ have been used  in the last equality.   
Since 
the effective Lagrangian is rather standard, this relation has been obtained previously e.g. 
\cite{Ellis:1970yd,Leung:1989hw,Cata:2018wzl}.
What is new is the result $\gast=1$ ($\Deqq=2$), 
that there are no $\bestp$-terms  and the derivation in itself from the double-soft theorem.

The  dilaton GMOR-relation quoted in \eqref{eq:D-GMOR} can be obtained by going through the analogous process 
for the pion which has been done in \cite{Zwicky:2023bzk}  but we shall repeat it here for completeness. 
Starting with  the analogue of \eqref{eq:start} one gets
\begin{alignat}{2}
 & 2 m_\pi^2 &\;=\;& \matel{\pi^a}{\TEMT}{\pi^a} =  \matel{\pi^a}{\Opqq}{\pi^a} =  \frac{-(1+\ga_*) \mq}{F_\pi} 
  \matel{0}{ i [ Q_5^a , \bar q \mathbb{1}_{\Nf} q]}{\pi^a}     \nonumber \\[0.1cm]
  & &\;=\;&  \frac{ 2 m_q (1+\ga_*)}{F_\pi} \matel{0}{P^a}{\pi^a}  =  \frac{- 2 \mq(1+\gast)}{F_\pi^2}  \vev{\bar qq}  \;,
\end{alignat}
upon using  
$\matel{\pi^a}{P^b}{0} =  - \frac{1}{F_\pi}  \vev{i [Q_5^a,P^b]}= - \frac{1}{F_\pi}  \vev{\bar qq}\de^{ab}$  
with $P^b$ as in \eqref{eq:OpsGB}. Rewriting 
\begin{equation}
\label{eq:GMOR}
m_\pi^2 F_\pi^2 =   - (1+\ga_*) m_q \vev{\bar qq} = - 2 m_q \qq \;,
\end{equation}
 the GMOR-relation  emerges \cite{Gell-Mann:1968hlm,Scherer:2012xha} upon using $\gast =1$. 
 Combining \eqref{eq:D-GMORa} and \eqref{eq:GMOR} 
  \begin{equation}
 \label{eq:D-GMOR2}
  F_D^2 m_D^2 = \Deqq(d-\Deqq)  \frac{\Nf}{2} F_\pi^2 m_\pi^2 \;,
 \end{equation}
 we get the dilaton GMOR-relation  \eqref{eq:D-GMOR}. 
 There is a further insight hidden here. Namely the $d$-term, which originated from the derivative term in \eqref{eq:partial}, 
in fact corresponds to the Zumino-type term in \eqref{eq:Zumino} (to see that one needs to expand to second order in $D$).  
This underlines the necessity of the integration by parts procedure applied once more.   It is satisfactory and important 
that full consistency with the EFT is attained. 
\item [d)]   Under item b) we learned that the soft theorem demands $\De_\Op = 2$. 
It is more often than not inferred  from the soft theorem that the field strength tensor squared  $\TEMT =   \be/(2g) G^2 $
generates a dilaton mass  in the chiral limit. 
However, this is not done from $\matel{D}{G^2}{D}$ but from $\matel{D}{G^2}{0}$ which gives the relation
\begin{equation}
\label{eq:PCDC}
  m_D^2 F_D^2 =  - \frac{2 \be}{g} \vev{G^2} \;, 
\end{equation}
upon using $\De_{G^2}=4$, 
 sometimes referred to as the  partially conserved dilatation current (PCDC) relation. 
Since $\De_{G^2} = 4$ does not meet the condition \eqref{eq:DeOp} this raises a question mark over the procedure. 
Either results, \eqref{eq:result1} and \eqref{eq:result2}, would give zero since $d-\De_{G^2} \to 0$. 
One may wonder whether this indicates that the gluon condensate is to vanish.
 A logical possibility that suggests itself is that 
 the dilaton could be massless in the chiral limit $\mbase \to 0$ and   
 that all three matrix elements vanish, 
$\matel{D}{G^2}{D} = \matel{D}{G^2}{0}  = \vev{G^2} =0$. We consider it worthwhile to emphasise this possibility 
without insisting on it (cf. the discussion in \SEC\ref{sec:softper}). 
\end{itemize}
In summary we have learned that the dilaton can only obtain a soft mass from an operator of 
scaling dimension two such as $\bar qq$ in the gauge theory.  
In addition, the soft theorems reproduce the  dilaton GMOR-relation \eqref{eq:D-GMOR2}. 
Single-soft theorems are equivalent to results found in more formal considerations e.g. \cite{Karananas:2017zrg}. 
It could be interesting to extend the techniques of both approaches to each other.

\section{Massive or Massless Dilaton?}
\label{sec:massless}
 
It is commonly believed that there are CFTs for which conformal symmetry is 
spontaneously broken, leading to a massless dilaton  and massive states  e.g. 
\cite{Schwimmer:2010za}.
The situation of when there is an RG flow into an IR fixed point cannot be 
regarded as settled.
The problem is that the symmetry is only emergent and we are not aware 
of a systematic treatment of this case.  Investigations in holographic approaches 
argue for a light but not a parametrically light dilaton   \cite{Coradeschi:2013gda,Belyaev:2019ybr,Pomarol:2019aae,Pomarol:2023xcc}.
However, since the answer to the question might be model-dependent 
 we  focus on our framework.\footnote{There are examples of massless dilatons 
in lower dimensions e.g.  $d=2$ \cite{Semenoff:2018yrt} at finite temperature and 
 $d=3$ \cite{Bardeen:1983rv} (cf. also \cite{Litim:2017cnl,Semenoff:2024prf}) but they do not involve an RG flow.  
 An example with a flow is given by a Gross-Neveu-Yukawa theory in $d=3$ where spontaneous scale symmetry,  
 emerges for certain initial conditions \cite{Cresswell-Hogg:2023hdg,Semenoff:2024prf} (cf. also \cite{Cresswell-Hogg:2022lgg,Cresswell-Hogg:2022lez} for related work), accompanied 
 with a massless scalar and massive fermions.  
 Explicit studies with fundamental scalars indicate  
 that they cannot take on the role of a dilaton \cite{Gildener:1976ih,Antipin:2011aa,Nogradi:2021zqw} 
 although they share some of these features. In \cite{Cata:2018wzl}  it was stressed that  
 scalars, called scalons in \cite{Gildener:1976ih}, are not to be regarded as dilatons.}
Parametrically,   the leading effect is expected to come from  the slope of the $\be$-function, 
which is however zero in our framework \eqref{eq:bestp} and thus  the parametric expectation moves to
\begin{equation}
m^2_D \propto \ORD(\bestpp) \;.
\end{equation}
This finding  can be taken as an indication  that the dilaton mass could at least be small. 
We are going to reflect on the question from three different points of view: 
the soft theorems, the \LNc limit  and the EFT-perspective. 
Whereas not conclusive, we hope that the reader finds the discussion  instructive.

\subsection{Soft-theorem perspective}
\label{sec:softper}

The form of the TEMT \eqref{eq:TEMTphys} is correct to all orders in perturbation theory 
and believed to hold beyond. If we set $\mq =0$ then there is only $\TEMT = \frac{\be}{2 g} G^2$
and thus we have $2 m_D^2 = \matel{D}{\frac{\be}{2 g} G^2}{D}$ in line with \eqref{eq:start}.
However, the double-soft dilaton theorem indicates that  the  dilaton mass ought to 
originate from an operator $\Op \subset \TEMT $ of scaling dimension $\De_{\Op}=2$ 
a result underlined by an alternative derivation from scaling in \APP\ref{app:massOp}.
Hence, this role 
cannot be taken by $G^2$ since  $\De_{\TEMT} = \De_{G^2} = 4 + \bestp=4$.\footnote{  
As stated earlier this puts into question the use of the  PCDC-type relation 
$m_D^2 F_D^2 = -2  \be/g \vev{G^2}$ \eqref{eq:PCDC}. 
To the best of our knowledge the relationship of  $\vev{G^2}$ in \eqref{eq:PCDC} to the gluon condensate introduced in phenomenology 
 \cite{Novikov:1977dq,SVZ79I} has never been clarified.
The latter has been determined empirically  \cite{SVZ79II,Ioffe:2002ee} and its existence is underlined 
by an elegant renormalon analysis within perturbation theory  \cite{Mueller:1992xz}.  
A similar quantity has been studied 
for pure SU(3) Yang Mills on the lattice  \cite{Bali:2014fea,Bali:2014sja,Bali:2015cxa} and is found to be non-zero by eleven standard deviation in the lattice scheme. The analysis is consistent with the
renormalon picture.}
As there is no other operator than $G^2$ present in the EMT (for $\mq =0$) this would then seem to imply 
that the dilaton is massless.\footnote{It is in principle conceivable that another operator becomes relevant 
and it its scaling dimension was $\De=2$ then this could give rise to a mass. We consider this possibility rather unlikely as
lattice studies of four fermi operators for example do not indicate large anomalous dimensions \cite{DelDebbio:2013uaa}.}     
We would not want to go as far as stating that this proves that the dilaton is massless,  
as for example there are  assumptions involved such as the use of the soft dilaton formula \eqref{eq:soft2}.  
The near conformal scaling dimension of large charge operators (on a cylinder) is dependent on both $
\ga_*$ and $\De$ and might be give rise to additional information \cite{Bersini:2024twu}.

\subsection{Large-$N_c$ consideration}
\label{sec:largeNc}

The large-$N_c$ limit  \cite{tHooft:1973alw,Witten:1979kh,Donoghue:1992dd} is a  useful tool for 
 QCD as it leads to simplifications  \cite{Lucini:2012gg,Donoghue:1992dd}. 
The following relations between two-point functions, with notation as in \EQs\eqref{eq:OpsGB} and \eqref{eq:Ops}, 
\begin{alignat}{2}
\label{eq:largNc}
 & \vev{\OS(x) \OS(0) }_c   &\;=\;&  \frac{2}{N_f} \vev{\OSa(x) \OSa(0) }(1 + \ORD(1/\Nc)) \propto \ORD(\Nc)   \;, \nonumber  \\[0.1cm]
& \vev{\OP(x) \OP(0) }   &\;=\;&  \frac{2}{N_f} \vev{\OPa(x) \OPa(0) }(1 + \ORD(1/\Nc)) \propto \ORD(\Nc) \;,
\end{alignat}
must hold since they are  distinguished by large-$N_c$ suppressed quark-disconnected diagrams (i.e. 
connected by gluons only). 
The factor $2/N_f$ takes into account the normalisation $\Tr[T^a T^b] =  \de^{ab}/2$ and 
 the subscript $c$  stands for the connected part  (and serves to remove  
  $ \vev{\OS(x) \OS(0) } \supset \vev{\bar qq}^2 \propto \ORD(\Nc^2)$ which is 
peculiar to the vacuum quantum numbers of $S$). 

 The leading graphs in \eqref{eq:largNc} are the connected planar ones of $\ORD(\Nc)$ as non-planar ones are
 $\ORD(g^4) \propto \ORD(1/\Nc^2)$ suppressed  since $g^2 \Nc$ is held fixed.  
 We point out that  quark-disconnected graphs  arising from $S$ and $P$ are of 
 $\ORD(\Nc^0)$  and falls in between the $N_c$-counting above. 
 
 Famously, one can deduce in this way that the $\eta'$ mass must go to zero in the \LNc limit. 
 In the deep-IR  the $\OPa$-correlator is dominated by the pion
 \begin{equation}
 \label{eq:OPa-x}
 \vev{\OPa(x) \OPa(0) }  \propto  \frac{ | \matel{0}{P^a}{\pi^a}|^2}{x^2}  \propto \ORD(\Nc) \;,
 \end{equation}
and the scaling is deduced from the  soft pion theorem
 $| \matel{0}{P^a}{\pi^a}|^2 \propto \qq/F_\pi \propto \ORD(\Nc^{1/2})$ with 
  $\qq \propto \ORD(\Nc)$ and $F_\pi^2 \propto \ORD(\Nc)$   \cite{Donoghue:1991qv,Scherer:2012xha}.
  Since it is part of the leading contribution its
   behaviour must be mirrored by the $P$-correlator and
  one concludes that $m_{\eta'} \to m_\pi =0$ and $F_\pi \to F_{\eta'}$  for $\Nc \to \infty$. 
   
  Does the same thing happen in the scalar channel? If the answer is yes, then  
  a massless dilaton would imply that  the corresponding flavoured states would  approach 
  zero in the \LNc limit. This is however in contradiction with a 
  lattice $SU(\Nc)$-study where the $a_0$-meson, the $0^{++}$-cousin of the pion, shows no sign of becoming massless 
  for increasing $N_c$  (cf. \cite{Bali:2013kia} \SEC 3.5 and also \FIG 14 in \cite{Lucini:2012gg}).  
  The only caveat is that these studies are performed in the quenched approximation but one cannot expect 
  these qualitative features to be overturned by unquenching.  We therefore conclude that either the dilaton 
  cannot be massless or that the dilaton must be subleading in the  connected $S$-correlator. 
  We will argue for the latter.
  
  In clarifying at what order the dilaton appears in the \LNc counting we consider the analogue of 
  \eqref{eq:OPa-x} in the deep-IR. Using the soft theorem \eqref{eq:match2}, 
   \begin{equation}
 \label{eq:OSa-x}
 \vev{\OS(x) \OS(0) }  \propto  \frac{ | \matel{0}{S}{D}|^2}{x^2}  \propto  \frac{\qq^2}{F_D^2} \frac{1}{x^2} \;.
 \end{equation}
 we learn that the scaling is hidden in  $F_D$ and as the latter is defined from 
  the coupling to the EMT \eqref{eq:FDTmunu}, we are lead 
to  consider the EMT-correlator 
\begin{equation}
\vev{T_{\mu \nu}(x)  T_{\rho \la}(0)} = t^{(0)}_{\mu \nu \rho \la} \, \Ga^{(0)}_{TT}(x^2) + 
t^{(2)}_{\mu \nu \rho \la} \, \Ga^{(2)}_{TT}(x^2) \;.
\end{equation}
The tensor structures   $ t^{(0,2)}_{\mu \nu \rho \la}$ correspond to spin 0 and 2 respectively 
and are dependent on $x_\al$ and $\mink_{\al\be}$.    
 Since the dilaton is of spin $0$ we have
   \begin{equation}
  \label{eq:cTT1}
  \Ga^{(0)}_{TT}(x^2)  \propto \frac{F_D^2}{x^2}    \;,
  \end{equation}
  and the \LNc behaviour of $F_D$ follows from corresponding behaviour of the correlator.
  To infer this we must have a look at the 
   EMT  of a non-abelian gauge theory which assumes the form 
\begin{equation}
T_{\mu\nu} =  \left( \frac{1}{4} g_{\mu\nu} G^2 - G_{\mu\la} G^{\la}_\nu \right) + 
\frac{i}{4}  \bar q \left(  \ga_{\{\mu } \Dra_{\nu \}} -  \ga_{\{\mu } \Dla_{\nu \}}   \right)  q + \dots \;,
\end{equation}
where  $\Dra_{\nu} =( \overset{\rightarrow}{\partial}  + i g A)_\nu$, $\Dla_{\nu} = (\overset{\leftarrow}{\partial} - i g A)_\nu$ and the dots stand for terms which vanish on physical states.  
The main point is that the gluonic part is in the adjoint and the quark part in the fundamental 
representation of the $SU(\Nc)$ gauge group. Hence, one  expects 
\begin{equation}
\label{eq:cTT2}
\Ga_{TT}  = A N_c^2 + B N_c + \ORD(\Nc^0)\;,
\end{equation}
and matching with \eqref{eq:cTT1} one infers that  generically  
\begin{equation}
F_D^ 2= a N_c^2 + b N_c + \ORD(\Nc^0) \;,
\end{equation}
is expected, implying 
$F_D \propto \ORD(\Nc)$.\footnote{This scaling is identified with glueballs as opposed to $F \propto \sqrt{N_c}$ 
which are referred to as $\bar qq$-states. 
In the literature one can find 
$F_D \propto N_c$ \cite{Golterman:2016lsd} and   $F_D \propto \sqrt{N_c}$ \cite{Crewther:2012wd}. We agree with the former reference who use the same argument without making it explicit.}
With \eqref{eq:OSa-x} it follows that the dilaton contribution is subleading in the \LNc limit
  \begin{equation}
 \label{eq:OSa-x2}
 \vev{\OS(x) \OS(0) }   \propto 
  \frac{\ORD(\Nc^0)}{x^2}  \;.
 \end{equation}
We therefore conclude that 
a finite value in $m_{a_0}$ at large $\Nc$ does not exclude a massless dilaton. 

 \subsection{Is dilaton-\chiPT consistent the existence of a massless dilaton?}
 
 Another way to assess whether a dilaton could be massless is to seek for 
 contradictions with the EFT.  
 Are quantum fluctuations going to induce a mass term? 
 In the chiral limit of  \chiPT\!\!  it is simply 
 impossible to write down a potential for the pion respecting the symmetries  
 within the coset construction $U$ \eqref{eq:Uchi} (due to the shift symmetry of the Goldstone).
 Hence the zero pion mass is built into 
 \chiPT naturally since chiral symmetry is present at all scales.  Returning to the dilaton we may 
 observe that  since there is no scale in the LO Lagrangian, in the chiral limit, no dilaton mass 
 can be generated either. 
 This conclusion is however too quick since scales enter through hadron masses.  
 The dilaton couples to  the nucleon mass term, e.g.  \cite{DelDebbio:2021xwu},
 \begin{equation}
 \label{eq:nucleon}
 \de \Lag_{m_N} = - \hat{\chi} m_N \bar N N \;.
 \end{equation}
 Generically such a term will  induce a dilaton mass term (e.g.  a single nucleon loop).   
 There are several loopholes in this argument. 
 First of all  nucleon chiral perturbation theory \cite{Gasser:1987rb,Becher:1999he,Scherer:2012xha} 
 is designed to compute nucleon properties due to a light pion cloud and not the other way around. 
 Whereas the nucleon is the only other stable hadron made out of light quarks 
 there are of course many other resonances such as the $\rho, \omega, \dots$ with light quark content.
 That opens the door to potential cancellations. This does in fact happen since bosons and fermions 
 contribute with opposite sign as exploited in supersymmetry and  the Veltman condition for the Higgs mass \cite{Veltman:1980mj}.  Moreover, in the case where a CFT is spontaneously broken, 
 e.g. \cite{Schwimmer:2010za}, the same problems would be apparent, but only apparent, as the dilaton is 
 believed to be a true massless Goldstone in that case. 
 Hence, one has to conclude that the EFT is not capable of making a definite statement about the 
 dilaton mass due to (potential) hidden cancellations.

 \section{The Dilaton Candidate in QCD: $\sig \equiv f_0(500)$}
 \label{sec:QCD}
 
 Let us turn to the dilaton candidate in QCD, the $\sig$-meson or $f_0(500)$  by  official PDG-terminology 
 \cite{PDG22}.  We replace  $D \to \sig$ honouring the name used by many particle physicists.  
 The $\sig$-meson has captured the interest and imagination of particle physicists for  long 
 as testified by its  history  and properties in a dedicated physics reports \cite{Pelaez:2015qba}:  
it is very broad, it does not fit well into a nonet structure, 
it defies Regge trajectories as well as qualitative  aspects of the  \LNc limit  \cite{Pelaez:2015qba}.\footnote{By 
nonet one means the  union  of the $SU(3)_F$-octet and -singlet which mix when $SU(3)_F$ is 
broken by non-degenerate quark masses.}  
The goal of this section is to apply \dchiPT at LO and try to see whether one can understand its mass 
and width semi-quantitatively. 
 
 The meaning of  
the mass and the width are given by  its pole    $\sqrt{{s_{\sig}}}= m_{\sig} - \frac{i}{2} \Gamma_{\sig} $, 
on the  second Riemann sheet,  in $\pi\pi$-scattering. Its current PDG-value \cite{PDG22} is 
\begin{equation}
\label{eq:sigpole}
\sqrt{s_{\sig}}=    (400-550)  - i (200- 350)  \MeV \;.
\end{equation}
This range is noticeably larger than  the specific determination 
 from  Roy equations $\sqrt{{s_{\sig}}} =  (441^{+16}_{-8} -i272 ^{+9}_{-12.5})   \MeV$   \cite{Caprini:2005zr},  which   is considered 
to have settled the issue of its existence.  Earlier determinations were compatible within larger uncertainties e.g. \cite{Oller:1998hw}.  It is often noted with regard to the $\sig$ being a pseudo Goldstone that its mass is notably heavier than that of the pion. However, this is not the 
right comparison since  the $\sigma$, just like the $\eta$, is an $SU(3)_F$-singlet for degenerate quark masses and thus retains sensitivity to the strange quark mass.  This can be seen in the dilaton GMOR-relation \eqref{eq:D-GMOR}. 

As  singlet-octet mixing will be relevant for the width,  
it is instructive to discuss the nonet structure and comparing it with the 
 $\rho$-meson family (cf. \TAB\ref{tab:nonet}).
 The qualitative differences  are apparent:
i)  the $u\bar d$-mesons ($I=1$) are lighter than the $u \bar s$-mesons ($I=1/2$) for  the vectors but heavier for 
the scalars, 
ii) the ratio of the $I=1$ octet to $I=0$ singlet is roughly one for the vectors  but a factor of two for the 
scalars.   These aspects challenge the quark model picture and can be seen as one the motivations 
for introducing the phenomenologically successful  tetraquark model  \cite{Jaffe:1976ig}.  
The widths also follow interesting patterns. 
The decays $a_0(968),f_0(980) \to \pi\pi$ are suppressed by $G$-parity and being mostly an $\bar ss$-state respectively. 
This is analogous to  the $\omega(780)$ and the $\phi(1200)$-meson in the vector channel. 
The $\kappa$ is the  $K^*(895)$-analogue  and indeed rather broad $\Ga_{\kappa} \approx 600(80)\MeV$, 
an aspect which we will understand better when considering the mixing in the next section.

We now turn to the decay constant $F_\sig$. 
For the unstable $\sigma$-meson, the matrix element \eqref{eq:FDTmunu} is not well-defined. 
However, the residue at the complex pole, which is generally complex, is well-defined and accessible 
via proper analytic continuation. 
It has been extracted in QCD through   $\matel{\pi}{\bar qq}{\pi}$ and 
$\matel{\pi}{\TEMT}{\pi}$ form factor input up to $\ORD(q^2)$ \cite{Moussallam:2011zg}.  
It is not straightforward to interpret this result in the context of this paper 
as it requires to understand the meaning of an unstable pseudo Goldstone and how this affects its representation.
The same remark applies to the $g_{\sig NN}$-residue extracted from $\pi\pi \to NN$ scattering  \cite{Hoferichter:2023mgy}.
The clarification thereof seems important   and we hope to return to this question in a future publication. 
 We may get an indirect  estimate from relation $g_{\sig NN}   =  \frac{m_N}{F_\sig}$ e.g \cite{DelDebbio:2021xwu} (akin to the Goldberger-Treiman relation 
$g_{\pi NN}   =  \frac{m_N}{F_\pi}$). In nuclear physics there are approaches  using 
$g_{\sig NN}$ in their LO Lagrangian such as the one-boson exchange model  
describing nucleon-nucleon scattering \cite{CalleCordon:2009pit}. 
The $\sig$-meson, amongst $\pi$, $\rho$ and $\omega$, describe the scattering phase shifts reasonably 
well.  From  the range in \TAB III  \cite{CalleCordon:2009pit} one infers  $g_{\sig NN} = 10(2)$ 
(cf. \cite{Wu:2023uva} for compatible results and also the fact that the real part in \cite{Hoferichter:2023mgy} is in agreement as well)
as a reasonable estimate.  
Remarkably,  using $m_N  =  0.93\GeV$,
\begin{equation}
\label{eq:Fsig}
F_\sig  =  \frac{m_N}{g_{\sig NN}} \approx  93(19) \MeV \;,
\end{equation}
a value close to $F_\pi = 93 \MeV$ emerges.  However, 
one ought to be cautious as this is a very difficult subject where systematics are difficult to estimate.

\begin{table}[t]
\addtolength{\arraycolsep}{3pt}
\renewcommand{\arraystretch}{1.3}
$$
\begin{array}{ l | l || lll}
J^{PC} \backslash  I & 0 & 0 & 1/2  & 1 \\ \hline
0^{++} & \sig \equiv f_0(500) & f_0(980) & \kappa \equiv K^*_0(700)   & a_0(980)  \\
\Ga_{0^{++}}    &    550(150) \MeV & 55(15)\MeV & 600(80) \MeV  &  90(50)\MeV   \\   \hline
1^{--} & \omega(780) & \phi(1020) &  K^*(895)   & \rho(770) \\
\Ga_{1^{--}}    &    6.86(13) \MeV &  0.016 \MeV & 52(12) \MeV  &  145(3)\MeV 
\end{array}
$$\
\caption[]{\small Nonet of $SU(3)_F$ which illustrates the special character of the $J^{PC} =0^{++}$ mesons
versus the more familiar and understood $J^{PC} =1^{--}$  vector mesons, given  
 in PDG-notation \cite{PDG22}. The double bar separates the singlet from the octet states in the $SU(3)_F$ limit. It seems relevant to mention that higher $f_0$-resonances all have considerably smaller widths than the 
$\sigma$-meson: $\Ga_{f_0(1370)} = 350(150)\MeV$, $\Ga_{f_0(1500)} = 108(33)\MeV$, 
$\Ga_{f_0(1710)} = 150(12)\MeV$ and $\Ga_{f_0(2020)} = 180(60)\MeV$  \cite{PDG22}.}
\label{tab:nonet}
\end{table}

 \subsection{The width of the  $\sig$-meson}
\label{sec:Fsig}

The width of the $\sig$-meson has been one of the early qualitative successes of the dilaton approach.  
It  is well approximated by $\Ga_{\sig} \approx \Ga_{\sig \to \pi\pi}$ since 
the photon  channel 
$\Ga_{\sig \to \ga\ga} = 1.7(4) \keV$ \cite{Hoferichter:2011wk} is highly suppressed as one would expect. 
 Its interest lays more in the possibility to learn about the $\sig$-meson  substructure  
\cite{Pelaez:2015qba}.  The amplitude into two pions  is described the effective coupling, $\Lageff = \frac{1}{2} g_{\sig \pi\pi} \sig \pi^a \pi^a $, 
which we can read off from the Lagrangian \eqref{eq:LO}\footnote{In an EFT framework (for off-shell $\sig$ e.g. \cite{Zwicky:2023fay}) $m_\sig^2 \to q^2$ where $q^2$ is the momentum entering the $\sig$-field.}
\begin{equation}
\label{eq:gsigpipi}
g_{\sig\pi\pi} =  \frac{1}{F_\sig}(  m_\sig^2 + (1-\gast) m_\pi^2 + \ORD(\bestpp,\mq^2))  \to \frac{m_\sig^2}{F_\sig}    (1+ \ORD(\bestpp,\mq^2)) 
\;,
\end{equation}
and  resembles earlier expressions 
\cite{Zumino:1970tu,Ellis:1970yd,Crewther:2015dpa}.  Differences are that $\gast=1$ is an open parameter and that 
$\bestp$-corrections with unknown coefficients are parameterised e.g. 
\cite{Crewther:2015dpa}.
The rate 
\begin{equation} 
 \Ga_{\sig\to \pi\pi}|_{SU(2)} =  \frac{3 |g_{\sig\pi\pi}|^2}{32 \pi m_{\sig}} \sqrt{1-4 \hat{m}_\pi^2}   \;,
\end{equation}
 follows from the 
$1 \to 2$ decay $d \Ga = \sum_{\pi\pi}  \frac{|{\cal A}_{\sig \to \pi\pi}|^2}{32 \pi^2 }\frac{|\vec{p}\,|}{m^2_\sig}  d \Omega $  \cite{PDG22} 
where  $ \frac{|\vec{p}\,|}{m_\sig}  =  \frac{1}{2}\sqrt{1-4 \hat{m}_\pi^2}$ is the velocity in the frame of the $\sig$-meson, 
$\int d \Omega \to  2\pi$ as the pions are identical particles and the factor of   $3$ results from the three pion  
channels.  The case with and without mixing are denoted by $SU(3)_F$ and $SU(2)_F$ respectively.

\paragraph{No mixing:}
using $m_\sig = 440\MeV$ and  $F_\sig = 93\MeV$ one gets  
\begin{equation} 
 \Ga_{\sig\to \pi\pi}|_{SU(2)} = 227 \MeV 
\end{equation}
where  uncertainties are not given since the mixing is neglected.   
The rate is a factor of $\approx 2.5$  lower  compared to \eqref{eq:sigpole}, which  corresponds  
to a factor  $\approx 1.5$ in the amplitude and not a bad results in view of 
the crudeness of 
the approach. The question is whether the singlet-octet refinement  will improve or worsen it. 

\paragraph{Singlet-octet mixing:}
the effect of singlet-octet mixing is driven by the breaking of $SU(3)_F$,  i.e.  $m_s \gg m_{u,d}$. 
It is known that this effect is not negligible  from the $\eta$-$\eta'$ system. In the same way the
$\sig$-$f_0(980)$ system may be parameterised by  a single angle $\theta$
\begin{alignat}{2}
\label{eq:18}
& \state{\sig} &\;=\;&  \phantom{- }\,\cos \theta \state{S_1}  + \sin \theta \state{S_8}   \nonumber \;,  \\
& \state{f_0(980)} &\;=\;& - \sin \theta \state{S_1}  + \cos \theta \state{S_8}  \;,
\end{alignat}
in terms of the  $SU(3)_F$ eigenstates. 
The isospin breaking mixing with the  $a_0$, the analogue of $\pi^0$ in the $\eta$-$\eta'$ system, is neglected.  
We follow the  approach by Oller \cite{Oller:2003vf} based on a   Wigner-Eckart decomposition  with two reduced matrix element $g_1$ and $g_8$ for which one has
\begin{equation}
{\cal A}_{\sig \to (\pi\pi)_0} = - \frac{\sqrt{3}}{4} \cos \theta g_1 - \sqrt{\frac{3}{10}} \sin \theta g_8 \;,
\end{equation}
 and similarly for $f_0(980)$ including all open channels. The zero subscript stands for $I=0$. 
The three unknowns $g_{1,8}$ and $\theta$ when fitted to experiment are
 (\EQ 2.30 \cite{Oller:2003vf})\footnote{There is a second determination $\theta  \approx 21^\circ$
  in an $U(3)\times U(3)$ $\sig$-model  \cite{Napsuciale:1998ip}. Whereas  no error is given in this determination,  presumably due to  model-dependence, 
 the agreement with \eqref{eq:Oller} is encouraging.}$^,$\footnote{We have used the singlet-octet to $\bar ss$-$\text{n}\bar ss$ basis conversion 
$ \theta =\phi +35.264^\circ $ for $\phi  \approx - 14^\circ$. I am grateful to 
Oller in assisting in the conversion 
which incidentally is not the same as in the standard $\eta'$-$\eta$ mixing.}    
\begin{equation}
\label{eq:Oller}
g_1 = 3.9(8) \GeV \;, \quad g_8 = 8.2(8) \GeV \;, \quad \theta  = 19(5)^\circ \;.
\end{equation}
 
We note the remarkable enhancement of the octet component  $g_8/g_1 \approx 2$. 
Its effect may be estimated by the the ratio of mixing versus no mixing
\begin{equation}
r_{18} = \left|\frac{{\cal A}_{\sig \to (\pi\pi)_0}^{\theta\phantom{=0}}}{{\cal A}_{\sig \to (\pi\pi)_0}^{\theta=0}}\right|  =  1.81\,^{-0.18}_{+0.20} \;.
\end{equation}
Using this value we get 
\begin{equation}
\label{eq:SU3}
  \Ga_{\sig \to \pi\pi}|_{SU(3)} =  r^2_{18} \Ga_{\sig\pi\pi}|_{SU(2)} =    744\,^{-108}_{+146}  \pm 40\% \MeV \;,  
\end{equation}
where  uncertainties were obtained by  adding the ones due to $g_1$ and $\theta$ in quadrature 
plus  $40\%$ for $F_\sig$.
Whereas the estimate is crude that is errors are large, it is fair to state that 
its central value is considerably improved due to the mixing and compares 
favourable with $\Ga_{\sig \to \pi\pi} \approx 544^{+18}_{-25}\MeV$ from the Roy equation  \cite{Caprini:2005zr}.
The enhanced octet versus singlet matrix element explains the large $\kappa$ rate 
(\TAB\ref{tab:nonet}). It is enhanced by $(g_8/g_1)^2$ and its natural value when compared with the $\sig$-meson 
is therefore  $140\MeV$  rather than $600\MeV$. It would be interesting to understand the octet enhancement 
qualitatively. 

 \subsection{The mass of the  $\sig$-meson}
\label{sec:msig}
 
In QCD the GMOR-type mass relation,  which are the   \chiPT\!\! LO expressions \cite{Donoghue:1992dd,Scherer:2012xha},
\begin{equation}
\label{eq:GMORSU3}
 m_\pi^2  =    (m_u + m_d) B_0 \;, \quad 
 m_{K^+}^2 =    (m_u+ m_s) B_0  \;, \quad m_{K^0}^2 =   (m_d+ m_s) B_0 \;,
\end{equation}
 work rather well for  pions and  kaons  (recall $B_0 = - \vev{\bar qq}/F_\pi^2$).   
 This raises hopes that  the dilaton GMOR-relation \eqref{eq:D-GMORa} will give a good for value of the $\sig$-mass. We shall see and understand that this is not necessarily the case.
 Let us first adapt \eqref{eq:D-GMORa} to non-degenerate quark flavours,   
 using \eqref{eq:GMORSU3}, and deduce
 \begin{equation}
\label{eq:msigSU3}
 F_\sig^2  m_\sig^2  =   (1 + \ga_*)(3 - \ga_*) F_\pi^2 \sum_{q=u,d,s} m_q    B_0  |_{\gast=1}
    = 2 F_\pi^2 (m_{K^0}^2 +m_{K^+}^2 + m_\pi^2)   \;.
\end{equation}
With $m_\pi = 140\MeV$, $m_K = 495 \MeV$, $F_\pi = 93\MeV$  and $F_\sig = 93\MeV$ as input 
on gets, 
\begin{equation}
\label{eq:msigSU3N}
m_\sig|_\LO  \approx 1 \GeV \left( \frac{ 93 \MeV}{F_\sig} \right) \;, 
\end{equation}
 a value which is about a  factor of two larger than the  in the real world.  
 There is some irony here as often the $\sig$-meson is regarded as being too heavy 
 to be considered a pseudo Goldstone.  Formula  \eqref{eq:msigSU3} has been obtained in 
 \cite{Crewther:2013vea}  but the difference is that $\gast$ and $\bestp$-corrections were undetermined 
 and thus the  large value has not  come to attention.
   
One might wonder whether \dchiPT would be convergent with such a large LO value 
since it is well-known that $SU(3)$-\chiPT is not 
as efficient as $SU(2)$-\chiPT because of the proximity of $m_K$ to  $m_\rho \approx 770\MeV$. 
The separation of the Goldstone sector is not strong in actual numbers. 
To get a an idea we might want to use the NLO formula for $SU(2)$-\chiPT  \cite{Scherer:2012xha,Leutwyler:2012}
\begin{equation}
m_\pi^2|_{\NLO} = m_\pi^2( 1- \frac{1}{32 \pi^2} \frac{ m_\pi^2}{ F_\pi^2} \bar \ell_3) \;,
\end{equation}
with $\bar \ell_3 \equiv \ln  {\La_3^2}/{m_\pi^2} = 3.53(26)$  \cite{FlavourLatticeAveragingGroupFLAG:2021npn} 
is sizeable due to a chiral logarithm.
One infers NLO-corrections factors of  
 $(0.025,0.21,0.25,1.4)$, using LO masses 
 $(m_\pi ,m_K,m_\sig, 1 \GeV  )$.\footnote{This procedure only gives realistic numbers for the pions 
 since the Kaon requires $SU(3)$-\chiPT  \cite{Gasser:1984gg}  and for the $\sig$  
 \dchiPT is required  which has not been fully developed at NLO yet. We believe however that the numbers 
 give a reasonable estimate of the size of the NLO corrections.} 
 The  large correction factor $1.4$ for $m_{\sig}|_{\LO}$ is telling us that   convergence cannot be expected.   The situation is unsatisfactory but deserves some more contemplation.

 
 The large value obtained is driven by $m_K$ which is comparatively 
 large due the  $m_s \gg m_{u,d}$. 
The  GMOR-type formulae are expansions in $\mq$ (or Goldstone masses) and not in $1/\mq$ . 
 That is, there is no built-in decoupling limit but rather one decouples by hand in excluding a quark  from
 the sum in \eqref{eq:msigSU3}. Can we assess this in another way?  
 Yes, through the mixing if we are willing to commit to a quark-mass picture. 
  In \APP\ref{app:mix} it was  argued that the mixing  angle supports the $\bar qq$-state interpretation.  
  \EQ\eqref{eq:qq} indicates suppression of the  $\bar ss$  versus the $\bar uu$,$\bar dd$-contribution. 
  The  strange quark decoupling angle   $\theta_{\text{dec}} \approx 35.7^\circ$ is close but not too close to 
  $\theta = 19^\circ$.\footnote{There is  indirect  evidence for  strangeness in the 
 $\sig$-meson.  The
   $\vev{\bar ss(x) \bar uu(0)}$-correlator is non-zero suggesting a light state which could correspond to 
     the   $\sig$-meson   \cite{Descotes-Genon:1999jcm}.} Indeed if we were to completely decouple the strange quark 
     we would expect to replace $2 m_K^2+ m_\pi^2 \to 2 m_\pi^2$  in \eqref{eq:msigSU3}  which does yield 
 $m_\sig \approx 2 m_\pi = 280 \MeV$   an underestimate.

In summary the situation remains inconclusive but we pointed out why 
we cannot expect formula  \eqref{eq:msigSU3} to give us a good number. 
First from the convergence in the EFT and second the formula overestimates the 
role of the strange quark since it has no decoupling built in.

 \section{Outlook}
 \label{sec:outlook}
 
 There are a few directions in which the work begun in this paper can be extended,  
 which is to investigate the $\mq$ dependence of the $\sig$-mass, extending the LO Lagrangian \eqref{eq:LO}
 to include $\mbase$-mass and investigate whether the Higgs boson could be a dilaton.  Below we give 
 a brief outlook on some of these matters.
 
 \subsection{The quark-mass dependence of the $\sig$-meson on the lattice and beyond}
 \label{sec:mqsig}
 
 Investigating the $\mq$-dependence of the $\sig$-meson in QCD  
 and other gauge theories with chiral symmetry breaking is interesting and still largely open.
 There are several avenues and it is to be hoped for efforts to continue and for new ones to start. 

 Before the advent of the  LHC  investigations in lattice 
gauge theory started  in order  to obtain reliable information for walking technicolor \cite{Hill:2002ap,Sannino:2009za,Piai:2010ma}
 and composite Higgs models \cite{Panico:2015jxa}.  
 In recent years  light scalars with $\sig$ quantum numbers were reported for decreasing quark masses
\cite{Hasenfratz:2020ess,Kuti:2022ldb,Fodor:2019ypi,Fodor:2017gtj,Fodor:2018uih,Fodor:2019vmw,Fodor:2020niv,Chiu:2018edw,LatticeStrongDynamics:2018hun,DelDebbio:2015byq,Brower:2015owo,LSD:2014nmn,LatKMI:2014xoh,LatKMI:2016xxi}. 
  There are also studies of the QCD $\sig$-meson on the  lattice  \cite{Briceno:2016mjc,Briceno:2017qmb,Hu:2016shf,Guo:2018zss} which is challenging because of its large width.   
As far as we know all simulations are performed at physical strange quark mass  but larger $m_{u,d}$  masses.  
Concretely, in \cite{Briceno:2016mjc,Briceno:2017qmb} a pion mass  $m_\pi  = 391 \MeV$ leads to   $m_{\sig} =745(5) \MeV $  and $m_{f_0(980)} =1166(45)  \MeV$. 
 Another interesting avenue is to measure the matrix element 
$\matel{0}{\bar qq}{D}$, from which the $\sigma$-mass is extracted.  
If the dilaton chiral mass is zero  then there is a simple relation with   $F_\sig$  of this matrix element  cf. \APP\ref{app:single}. 
In fact, the $SU(3)$ and $\Nf=8$ lattice results is compatible with a very light or possibly massless dilaton. 

Amongst the analytic methods the Roy equations  \cite{Ananthanarayan:2000ht,Caprini:2005zr}, 
which give the most precise determination of the $\sig$-pole,  would seem 
the most promising method. They do rely though, as many dispersive methods, on  
a mixture of experimental and theoretical hadronic input. Hence the success would depend 
on how well one could control the input as a function of the quark masses.  The lattice could play an indirect  but 
important role 
in providing input for the Roy equations. For example, recently the data from \cite{Briceno:2016mjc,Briceno:2017qmb} 
has been used in Roy equations \cite{Cao:2023ntr}  and the $m_{\sig} =745(5) \MeV$ consistent with \cite{Briceno:2016mjc,Briceno:2017qmb}  has been found. 
  Another possible avenue is the analytic $S$-matrix bootstrap 
\cite{Kruczenski:2022lot,Guerrieri:2020bto} which would equally depend on input. 
Analytic methods such as Dyson-Schwinger equation \cite{Santowsky:2020pwd,Santowsky:2021ugd} and Functional Renormalisation Group   
 Methods based on elastic unitarity, $\pi\pi$-scattering $s < 4 m_\pi^2$, have been applied 
 (reviewed in \cite{Pelaez:2015qba}). Concrete studies include the inverse amplitude method 
 with NLO \chiPT input for fixed $m_s$ and varying pion mass \cite{Hanhart:2008mx}, 
unitarised chiral perturbation theory  \cite{Nebreda:2010wv} with varying $m_s$-dependence 
and the $N/D$-method with LO \chiPT\!\!-input \cite{Oller:2003vf}.

The numbers from  lattice \cite{Briceno:2016mjc,Briceno:2017qmb,Rodas:2023twk} and analytic methods indicate that  $\sig$ could decrease by  $\ORD(100\MeV)$  
for in $m_{u,d} \to 0$.
Hence, the role of the strange quark mass might be important for which  information is sparse. 
The above mentioned $N/D$ method \cite{Oller:2003vf} gives most concrete 
information where the nonet, in mass and widths, are continuously deformed  
to become $SU(3)_F$-symmetric at $m_{\pi,K} = 350\MeV$ (corresponding to a degenerate quark mass $\approx 23 \MeV$,  
an increase in $m_{u,d}$ and a decrease in $m_s$).  
At this point $m_\sig \approx 300\MeV$ as can be inferred from 
the plot in \FIG 2 in \cite{Oller:2003vf}.
This is a significant reduction in view of the expected increase 
due to $m_{u,d}$.\footnote{Note that for small deformation  it is the width rather than the mass that decreases. 
This also explains why the $\sig$-mass shows little variation for small $m_s$ in the 
in \REF\cite{Nebreda:2010wv}.}  This is somewhat at variation with   \cite{Nebreda:2010wv} 
where (small)  $m_s$-changes  in mass and  width were found to be very small.  
Hence the situation is not entirely conclusive 
but rather motivates to further investigate $m_\sig$ as a function of the light quark masses.

 \subsection{Is the Higgs a dilaton?}

 It has been appreciated for a long time that 
in the absence of the Higgs VEV the SM is conformal up to the  logarithmic 
 running of the couplings. 
 The dilaton therefore fits the role of the Higgs naturally as it couples to mass and is associated
 with a VEV.  There are several realisations of this scenario but they all have in common 
 that the SM Higgs sector is replaced by a (strongly coupled) sector which undergoes 
 spontaneous scale and electroweak symmetry breaking at a scale $F_\sigH$ (not necessarily  
 equal to the Higgs VEV $v \approx 250\GeV$). 
 The most relevant change is that in  the LO-SM Lagrangian the dilaton replaces the Higgs   as follows 
 \begin{equation}
 \label{eq:compensator}
 1 + \frac{h}{v}  \to  e^{-\frac{\sigH}{F_\sigH}} \to  1 + \frac{h}{F_\sigH}  \;,
 \end{equation}  
 where in the second arrow the freedom of field 
redefinition has been made use of.\footnote{This is legitimate as long as we are interested in on-shell matrix elements 
and for small fluctuations ($h < -F_\sigH$ would violate the positivity of the exponential).} 
 This particle behaves like a SM  Higgs with all couplings rescaled by   
the ratio of the two scales: $r_v = \frac{v}{F_\sigH}$.  Post LHC we know that this ratio has to be close to one within 
approximately ten percent which would equally tame tensions 
with electroweak precision  observables  
(e.g. \cite{Pich:2013fea} where $r_v=1$ implies $\kappa_W =1$). 
 This raises the  question: is $r_v \approx 1$ natural?
  Other important low energy parameters are the dilaton mass and more generally its potential. 
The answer to all of these questions is model-dependent and we thus restrict  attention to our framework.

 
We consider a gauge theory with a gauge group $G'$ which undergoes chiral symmetry breaking 
close to the electroweak scale. This means that 
the $W$- and $Z$-boson masses are generated in the same way as in technicolor \cite{Hill:2002ap,Sannino:2009za,Piai:2010ma}.  
The Higgs VEV and the pion decay constant of the new sector are related by 
 $v^2  = n_d  F_\pi^2 $, where $n_d$ is the number of techniquark electroweak doublets. 
Since the Higgs width has been measured to be narrow 
 $\Ga_h = 3.2^{+2.8}_{-2.2}\MeV$ \cite{PDG22} there can only be one doublet, which 
 takes on the role of   the longitudinal degrees of freedom of the gauge bosons. Otherwise    
 the Higgs/dilaton  would disintegrate fast into the additional $\pi'\pi'$-pairs giving rise to a width   
 $\Ga_h  = \ORD(100\GeV)$. 
Hence, $v = F_\pi$ and therefore  the difference of the 
 dilaton  versus  SM-Higgs coupling are parameterised by the  ratio 
  \begin{equation}
  \label{eq:r}
 r_{N_f} = \frac{F_\pi}{F_D}  \;, \quad 
 \end{equation}
 of pion to dilaton decay constants. Besides the indicated $N_f$-dependence there is also an 
 implicit $G'$-dependence which is less important but has to be kept in mind.   
 Whereas it is likely that $r_{N_f} = \ORD(1)$ there is no known reason why it should be numerically close to one. 
 The quantity we are interested in is $ r_{2}$ and we  resort to actual QCD for guidance.  
 With $F_\sigH$ as in \eqref{eq:Fsig} and $F_\pi = 93 \MeV$ one finds 
a number  $r_{2-3}|_{\QCD}= 1.0(2)$ which is compatible with one within uncertainties. 
Cautionary remarks apply. First  it is difficult to estimate the systematics of \eqref{eq:Fsig}. 
Second whereas we are interested in $N_f =2$ massless quarks, in QCD we have two light quarks and 
a light but sizeable strange quark.\footnote{The $N_f$-dependence comes from  the $\sig$-meson 
being a singlet $SU(N_f)$  as manifested in \eqref{eq:D-GMOR}.  Hence on expects 
$r_{N_f} \propto 1/\sqrt{N_f}$.  In a lattice fit  \cite{LSD:2023uzj} 
$r_8  \approx 0.33$  found for which the scaling gives  $r_2 = 2 r_8 = 0.66$ a value  slightly lower than $1$. 
There could be many  reasons, one of which is that the \eqref{eq:D-GMOR} is only a LO relation 
and the analysis in \SEC\ref{sec:largeNc} indicates that the counting is not that straightforward.}
The most pragmatic interpretation  is that  this motivates trying to find a reason for $r_2$ being  close to one.
 
 So far we have not addressed how the coupling  
 \eqref{eq:compensator} arises.\footnote{In the most commonly used dilaton scenarios  
this follows from the conformality of the total SM and extended sector. This is not the route we have in mind 
as this scenario has  sizeable contributions in $gg \to h$ for example,  in tension with the  
findings at the LHC \cite{Serra:2013kga}.} 
The starting point is  the higgsless SM  developed as an EFT for a heavy Higgs 
 \cite{Appelquist:1980vg,Longhitano:1980iz,Buchalla:2012qq}.  This is analogous to decoupling the $\sig$-particle 
 in the linear $\sig$-model which gives a for of \chiPT\!\!.
 The would-be Goldstone bosons 
$U = \exp(i 2 T^a \pi^a/F_\pi)$ transform as $U \to V_L U V_Y$ under the SM gauge group 
$SU(2)_L \times U(1)_Y$ with $V_Y = e^{i y T_3}$ such that the condition $\det U=1$ is preserved (since $\det V_L = \det V_Y =1$). 
The effective Lagrangian contains terms of the form 
\begin{equation}
\label{eq:higgsless}
\Lag \supset  \frac{1}{4} v^2 \Tr [ D^\mu  U D_\mu  U^{\dagger}]  -  v \bar q_L Y_d U {\cal D}_R  + \dots  \;,
\end{equation}
where  $\bar {\cal D}_R \equiv (0,\bar d_R)$ and the covariant derivatives assure $SU(2)_L \times U(1)_Y$ gauge invariance.
The dots correspond to  similar fermionic  and gauge kinetic terms. 
\EQ\eqref{eq:higgsless} resembles the corresponding \dchiPT term in \eqref{eq:Lagkin} but lacks the dilaton nota bene. 
It seems to us that one cannot resort to the compensator argument in coupling the dilaton since it is only 
the scale symmetry of $G'$ which is broken. However, the same rationale that would imply  $r_2 \approx 1$ 
ought to group the pions and the dilaton under one and the same symmetry.  Assuming this to be true then gives
\begin{equation}
\label{eq:higgsless2}
\Lag \supset    \frac{1}{4} v^2 e^{-2 \sigH/F_\sigH}  \Tr [ D^\mu  U D_\mu  U^{\dagger}] -  v e^{-\sigH/F_\sigH} \bar q_L Y_d U {\cal D}_R + \dots  \;,
\end{equation}
an effective Lagrangian  equivalent to the  SM  at LO modulo the Higgs potential.  
Hence, finding an (approximate) symmetry assuring $r_2 \approx 1$  would therefore help in both ways, 
explaining the required closeness to the  SM and justifying the Lagrangian in \eqref{eq:higgsless2}.
Our reasoning  is the same as in   \cite{Cata:2018wzl}, except that we give provide a reason for dilaton interactions. 

Let us turn to the dilaton potential which consists of the pure $G'$-part and the one induced by 
the mixed Lagrangian \eqref{eq:higgsless2}.
For the former the situation is similar to \CDQCD, we do not know anything for certain  other than 
how to incorporate  $q'$ mass terms.  Now, since  $m_{q'} \neq 0$ 
breaks $SU(2)_L$, these terms are absent in the most straightforward setting.   
If the dilaton were to acquire a chiral mass  $\mbase \neq 0$ then 
$V_{\De}$ \eqref{eq:V} is a simple potential describing it.  However, since the dilaton soft theorems (in \SEC\ref{sec:soft})
indicate that $\De=2$ (which corresponds to the SM Higgs potential), it remains unclear which operator would take on this role
in the $G'$ gauge theory. 
 Let us turn to the potential induced by the mixed Lagrangian \eqref{eq:higgsless}.  
 As is the case in composite Higgs models \cite{Panico:2015jxa} the coupling to the $W,Z$-boson and the top quark would induce sizeable corrections to the Higgs mass which are quadratically sensitive to the cut-off $\La' = 4 \pi F_\sigH$ of the $G'$-confinement scale.  It is difficult to say anything concrete other than parametrically these contribution are of order 
 $\ORD(v^2)$ subject to further sizeable NLO corrections. 
In view of the generally large and negative contribution of the top mass 
a sizeable chiral  mass $\mbase$ could potentially be required. A more thorough assessment 
might necessitate to find a UV completion at some scale $M^2$, explaining the origin  of  the 
coupling \eqref{eq:higgsless2} through terms of the type   $\Lageff \supset \frac{\ORD(1)}{M^2} \bar q' q' \bar tt $.

There are further phenomenological aspects that needs attention. In the 
  standard dilaton scenarios the  
 $\be$-function contributions to the SM radiative processes $gg \to h$ and $h \to \ga\ga$ 
 give rise to severe constraints e.g. \cite{Serra:2013kga}. 
 In our scenario  the $q'$-fermions are not charged under  $SU(3)_c$, implying that  
 $gg \to h$ is  truly loop-suppressed (with respect to the SM) and therefore we do not expect  tensions with the LHC. 
The process $h \to \ga\ga$  
is more subtle as the $q'$-fermions are generally electrically charged  (since $q'_L$ is charged under $SU(2)_L$). 
Its calculation is a formidable task as  non-perturbative. Early  assessments within QCD go back to the 
discovery of the trace anomaly \cite{Crewther:1972kn,Chanowitz:1972vd,Chanowitz:1972da}.
This LO evaluation  gives the correct order of magnitude e.g. \cite{Crewther:2012wd}, but its precise value 
 is extracted indirectly from scattering data (e.g. \cite{Hoferichter:2011wk,Dai:2014zta}). 
Since there is no precise method of direct computation and the rate might well  be $m_\sig/m_N$-dependent,
it seems that one cannot easily borrow an estimate from QCD.  The QCD-value  $\Ga_{\sig \to \ga\ga} = 1.7(4) \keV$ indicates 
however that it could be sizeable and therefore deserves attention. 
There is another challenge and that is the non-discovery  of  hadronic $G'$-resonances.  
Whereas the dilaton itself might be light, the basic scale is set by the Higgs VEV $v= F_\pi \approx 250 \GeV$.  
A  $2 \TeV$-benchmark gives a ratio $2 \TeV /F_\pi|_{G'} = 8$ which is comparable to 
the one in QCD:  ${m_\rho/F_\pi}|_{\text{QCD}} \approx 8.3$.   Without a phenomenological analysis and 
a more precise assessment of the ratio it seems difficult to make a  statement other than
 that these resonances may  be within the future LHC-reach. 

In summary the by far most important task is to investigate whether there is  a reason for 
$r_2 \approx 1$ in \eqref{eq:r}. 
If this were true then our rough assessment indicates that the model might pass current constraints. 
We hope to return to these topics in future work.

 \section{Conclusions and Summary}
 \label{sec:conc}
 
 In this paper we further explored the possibility that gauge theories in the chirally  broken phase 
 admit an IR fixed point interpretation (cf. \FIG\ref{fig:CW}). 
 {In \SEC\ref{sec:SUSY} it was argued that this idea makes sense in  
  ${\cal N}=1$ supersymmetric  based on  
 the free meson  and the   squark-bilinear duality in the infrared-free magnetic and electric phase.}
 
 Scaling dimensions  were deduced by matching the gauge theory  fixed-point behaviour   
 with the effective (dilaton)-\chiPT (\SEC\ref{sec:deep}).
 The mass anomalous dimension at the fixed point is found to be $\gast =1$, 
 providing a more in-depth analysis of our earlier findings \cite{Zwicky:2023bzk}.  
 The vanishing    of  the $\be$- and $\ga_m$-slopes 
 at the fixed point: $\bestp = \gastp =0$, cf. \eqref{eq:bestp} and \eqref{eq:gastp}, constitute 
  new results. In particular, $\bestp =0$ has many attractive features: i) it supports a mass-gap interpretation,  
 ii) it is consistent with ${\cal N}=1$ supersymmetry (\SEC\ref{sec:SUSY})\footnote{Furthermore,  
 we argue that  $\gast =1$ in the chirally broken phase provides a correct description of ${\cal N}=1$ Seiberg-duality aspects.}
  and,  
 iii) it implies 
  logarithmic running of the  fixed-point coupling $\de g \equiv g - g^*$  \eqref{eq:log}. 
  The latter makes it plausible  that the trace anomaly is  reproduced by the generalisation 
 of the energy-momentum tensor for chiral theories (to include a dilaton).

 In the second part of the paper we focus on the dilaton $D$, the (pseudo) Goldstone boson due 
 to spontaneous scale symmetry breaking.  
 From  soft dilaton theorems in \SEC\ref{sec:soft} 
 it was deduced that a dilaton mass generating operator  $\Op \subset \TEMT$ is 
 necessarily of scaling dimension $\De_{\Op}=2$.
 An important role
 is played by the partial derivative  $x \cdot \partial$-term in the dilatation commutator  \eqref{eq:partial}: 
 i)   it renders the mass positive and ii) it is the counterpart of the Zumino-term \eqref{eq:Zumino} in  \dchiPT\!\!.  
These results are model-independent.  For $\mq =0$ there is no operator of scaling dimension two 
and this sets a question mark on the PCDC-type  relations \eqref{eq:PCDC}  based in $\De_{G^2}=4$. 
For $\mq \neq 0$ the ${\cal O} = m_q \bar qq$ takes on the role of $\De_{\Op}=3- \gast =2$  and might be seen as another reasoning 
for $\gast =1$. 
 In \SEC\ref{sec:massless}  we contemplate whether the dilaton could be massless or not.    
 We found that combining lattice data and $F_D \propto N_c$ does in principle still allow for a massless dilaton. 
 Reasoning in terms of spontaneously broken CFT suggests that the EFT entails subtle cancellations.

 In the third part     applications of the dilaton are explored. 
 In \SEC\ref{sec:QCD} we consider whether 
 the $\sig$-meson in QCD could be a dilaton.  Singlet-octet mixing  is found to be important as it considerably improves 
 the prediction of the width   \eqref{eq:SU3}.   
 The leading order $\sig$-mass, which follows from the dilaton GMOR-relation \eqref{eq:msigSU3}, is rather large
 and indicates convergence issues in the EFT (due to the large strange quark mass). 
 The  dilaton as the low energy part of a new gauge sector,  can take on the role of the Higgs boson 
 provided that  the ratio of pion to dilaton decay constants  $r_{N_f} \equiv F_\pi/F_D$ is unity for two flavours 
 (cf. \SEC\ref{sec:outlook}).   Whereas in QCD indications are that  $r_2 \approx 1$ 
  there is  no  underlying principle known why this ought to happen.  
 An assessment beyond leading order would give rise to a potential and corrections to  radiative processes such as $gg \to h$ and $h \to \ga\ga$  which  might pass current LHC-constraints.

There are many directions to explore but among the unresolved questions the most outstanding ones are:
 i) what is  the chiral dilaton (or $\sig$-meson) mass?  Is it zero or at least considerably smaller than the 
 nucleon mass and how does it depend on the distance to the lower boundary of the conformal window?
 ii)  is there a principle that would imply $r_2 \approx 1$? {iii) a more systematic investigation
 gauge theories away from the near-conformal regime (cf. \APP\ref{app:single} for motivation and one route)}
 These questions might well be very difficult  to answer by pure reasoning. 
 Hopefully lattice Monte Carlo simulations and  methods
 using analyticity and unitarity can be of some guidance.

\paragraph{Acknowledgments:}
During large periods of this work RZ was supported by a CERN associateship and an STFC Consolidated Grant, ST/P0000630/1.   
I am grateful to  Claude Amsler, Gunar Bali, Latham Boyle, Jorge Camalich, Gilberto Colangelo, Matthew McCullough, 
Poul Damgard, Luigi Del Debbio, Jo Dudek, Shoji Hashimoto, Martin Hoferichter, James Ingoldby, 
Andreas J\"uttner, Max Hansen, Georgios Karananas, 
Heiri Leutwyler, Daniel Litim, Tom  Melia, Jos\'e-Antonio Oller, Bashir Moussalam, Agostino Patella,
Jos\'e-Ramon Pelaez,  Werner Porod, G\'eraldine Servant, Misha Shifman, Christopher Smith, Yuji Tachikawa, Neil Turok 
and the participant of ``asymptotic safety meets particle physics" at DESY for useful discussions. 
At last apologies if I missed your work. 

\appendix

\section{Conventions}
\label{app:conv}

The Minkowski metric $\mink_{\mu\nu}$  reads  $\textrm{diag}(1,-1,-1,-1)$, $\gm = \det(\gm_{\mu\nu})$ is the determinant. 
Weyl transformations are defined by 
\begin{equation}
\label{eq:Weyl}
\gm_{\mu\nu} \to e^{-2 \al} \gm_{\mu\nu} \;,\quad \hat{D} \to \hat{D} -  \al \;,
\end{equation}
where the normalised dilaton field is $\hat{D} \equiv D/F_D$. 
Kinetic terms are shortened  as $(\partial \varphi)^2 \equiv  \partial_\mu \varphi \partial^\mu \varphi $.
The $\be$-function is defined in \eqref{eq:ga} and is given by 
\begin{equation}
\label{eq:be}
\frac{\be}{g} =  -  ( \be_0  \frac{g^2}{(4 \pi)^2} + \ORD(g^4) )    \;, \quad       \beta_0 = ( \frac{11}{3} C_A- \frac{4}{3} N_F T_F)  \;,
\end{equation}
where for QCD with $N_f =3$ and $G=SU(3)$ we have $\be_0 = 9$ since $T_F = \frac{1}{2}$ and $C_A = N_c$. Further to that we derive $\gaG$ in \eqref{eq:def}. It departs from the observation that 
$\TEMT = \frac{\be}{2g} G^2$ (for $\mq =0$) is an RG invariant 
\begin{equation}
0 = \frac{d}{d \ln \mu} \TEMT =   \frac{d}{d \ln \mu} \left( \frac{\be}{2g} G^2\right)  =  \left( \frac{\be}{2g} \right)'  \be G^2 +  
\frac{\be^2}{2g} (G^2)' \;,
\end{equation}
where the prime denotes differentiation with respect to $g$ and from  \eqref{eq:ga} one then 
gets
\begin{equation}
\gaG = 2g \left( \frac{\be}{2g} \right)' = \be' - \frac{\be}{g} \;,
\end{equation}
in accordance with \eqref{eq:def}.

\section{More on  Soft Theorems}
\label{app:soft}

In this appendix we take a look at the results  in \SEC\ref{sec:soft} from a slightly different angle. 
First in deriving  formula \eqref{eq:result2}  directly from \dchiPT\!\! and second by 
considering the  soft theorem for a single dilaton. 
The common theme is   the matching of  
\begin{equation}
\label{eq:Oqq}
\Opqq \equiv (1+ \gast) \sum_q \mq \bar qq  \;,
\end{equation}
which we can write as $\Opqq  = (1+ \gast) \partial_{\ln \mq} {\Lag}_{\QCD}$ in QCD
to  \dchiPT \eqref{eq:LO}
\begin{equation}
\label{eq:qqmatch}
 \Opqq \to (1+ \gast) \partial_{\ln \mq}   \LagLOdchiPT =
 (1+ \gast) \sum_q \mq \vev{\bar qq} \hat{\XX}^{\Deqq} =  -  \frac{1}{2} m_D^2 F_D^2 \, \hat{\XX}^{\Deqq}  \;,
\end{equation}
where the pions are neglected as they play no role in this appendix. 
In the last equality  \eqref{eq:GMOR}  was used and we note that since $\vev{\hat{\XX}^{\Deqq}} =1$ the VEV
$\vev{\Opqq} = - \frac{1}{2} m_D^2 F_D^2 $ is correctly reproduced. A peculiar aspect is that this operator contains tadpoles
\begin{equation}
\label{eq:expand}
\Opqq   =  -  \frac{1}{2} m_D^2 F_D^2 ( 1 - \Deqq \hat{D}  + \frac{1}{2} \De_{\bar qq}^2 \hat{D}^2 + \ORD(\hat{D}^3)  ) \;.
\end{equation}
We note that whereas in a Lagrangian tadpoles are not acceptable as they signal a false vacuum, there is nothing 
that prevent tadpoles in operator matching. There is another way to look at this since we state 
$\TEMT  = \Opqq $ the same must result from \dchiPT  \eqref{eq:LO}. Using 
$\TEMT = 4 V - \partial_{\ln \chi} V $ \cite{Zwicky:2023fay} we get $\TEMT = -\frac{1}{\Deqq} {m_D^2 F_D^2} \hat{\chi}^{\Deqq} $ 
which indeed matches provided  $\Deqq=2$ (or $\gast=1$).  In this matching we have used $d=4$ for simplicity. 

\subsection{The double-soft theorem from  dilaton-\chiPT}
\label{app:Lag}

We consider it important to make contact with the result    \eqref{eq:result2} directly from the Lagrangian. 
Whereas  in a general CFT setup  \eqref{eq:result2}  holds for any $\De_{\Op}$ this is not the case here as 
it was already concluded that $\De_{\Op} = d-2$. We aim to reproduce this result.

  Using \eqref{eq:expand}  and  \dchiPT  \eqref{eq:LO}  for the tadpoles we find 
\begin{alignat}{2}
\label{eq:cqe}
& { \matel{D}{\Opqq}{D} } &\;=\;&  \left( \matel{D} {  \frac{1}{2} \hat{D}^2 (  \De_{\bar qq}^2-  \Deqq(d+\Deqq) )  + 
 \frac{1}{2} \frac{( \partial \hat{D})^2}{m_D^2} \Deqq (d-2) }{D} \right) {\vev{\Opqq} } \nonumber \\[0.1cm]
 & &\;=\;& \frac{1}{F_D^2} ( \De_{\bar qq}^2-  \Deqq(d+\Deqq) + \Deqq(d-2))  {\vev{\Opqq} }  = 
-  \frac{1}{F_D^2}   2 \Deqq {\vev{\Opqq} } \;, 
 \end{alignat}
 for $\Opqq$ defined in \eqref{eq:Oqq} and subtleties due to tadpoles are
comment  further below. 
Using ${\Deqq=d-2}$ one finds 
indeed  consistency with  \EQs \eqref{eq:start} and  \eqref{eq:result2} and our goal is achieved. Namely,  we have shown that 
the soft theorem manipulations follow from the EFT provided that ${\Deqq=d-2}$. 

We consider it worthwhile to comment on the origin of the terms in \eqref{eq:cqe}. The   $\De_{\bar qq}^2$-term is straightforward as it corresponds 
to  the $D^2$-term in \eqref{eq:expand}.  The remaining two originate from tadpole diagrams 
due to the linear $\hat{D}$-term in  \eqref{eq:expand}.  For those the dilaton propagator assumes 
the form $\De_F(q^2=0) \to  -i/m_D^2$ and the mass term gets cancelled against a corresponding term in the 
 Lagrangian. 
Specifically, the  $\hat{D^2} \De(d+\Deqq)$- and $   (\partial \hat{D})^2  \Deqq (d-2)$-terms are due 
to the two interaction terms
\begin{equation}
 \LagLOdchiPT  \supset (d+ \Deqq) \, m_D^2 F_D^2   \frac{\hat{D}^3 }{3!}-  (d-2) \, F_D^2 \hat{D}
 \frac{( \partial \hat{D})^2}{2} \;.
\end{equation}

\subsection{The mass operator in the deep infrared}
\label{app:massOp}

We aim to derive  the result    \eqref{eq:DeOp}  using the the intuitive idea that the free field mass operator ${\Lag } \supset \frac{1}{2}m^2 \phi^2$ is of scaling dimension two. This is done by 
matching correlators in the full and in the effective theory in the deep-IR.   
We assume that the mass is generated by an operator $\la \Op \subset \TEMT$ where $\la$ serves as a bookkeeping parameter.  (If $\la = \mq$ then $\Op = \bar qq$ for example).
\begin{itemize}
\item \emph{EFT:}  the TEMT at LO assumes the form 
\begin{equation}
\label{eq:schematic}
\TEMT|^\LO_{d\chi\text{PT}} =   m_\pi^2 \pi^2  + F_D m_D^2 D -\frac{1}{2} \De_\Op m_D^2 D^2 \dots \;,
\end{equation}
where the dots stand for terms which give suppressed contributions 
in the deep-IR.  Assembling all the information one  gets
\begin{equation}
\label{eq:schematic2}
\CoE{ \TEMT}^\LO_{\text{d}\chi\text{PT}} \propto   \frac{  F_D^2   }{(x^2)^{d/2-1}}  e^{-m_D x} +
  \frac{1}{(x^2)^{d-2}}( c'_{2\pi} e^{- 2 m_\pi x}  +     c'_{2D} e^{- 2 m_D x} )     \;, 
\end{equation}
where  the exponential behaviour follows from the asymptotic limit of the Euclidean scalar
propagator $\De_E(x,m) \propto e^{-mx} (x^2)^{1-d/2}$.
\item \emph{RG analysis:} 
with  arguments similar to the ones in \SEC\ref{sec:RGder},  the  TEMT correlator assumes the form  
\begin{equation}
\CoE{\TEMT}_{\CDQCD} = \frac{\mu_0^{8}}{(\hat{x}^2)^{d}}   \, g(\hat{F} \hat{\la}^{-1/y_\la}
\hat{F} \hat{m}_q^{-1/y_m} , \hat{F}\hat{ x}^{d/2-1},  \mu) \;,
\end{equation}
with $y_\la  = d - \De_{\Op}$, $y_m = d-3 + \gast$, $g$ a dimensionless function and  $\mu_0$   some arbitrary reference scale used such that all hatted quantities are   dimensionless.  
Above we used   that $F_D$ is of mass dimension $d/2-1$ and 
$\De_{\TEMT} =d$ \eqref{eq:Dels} since $\bestp=0$.
  In order to regain 
 predictiveness we must know the $\la$- and the $F$-behaviour.  First, since the TEMT 
is proportional to $m_{\pi,D}^2 = \ORD(\la)$ it follows that $\CoE{\TEMT} = \ORD( \la^2)$ and thus the scaling 
exponent effectively  changes from $d \to  d - y_\la =  \De_{\Op}$ as one would expect.    
We may then drop the first argument in $g$ and focus on the $F$-dependence. 
For the $F$-dependence we need to think in terms of the EFT.  For the two Goldstone case there is no $F$-dependence 
\eqref{eq:schematic}
and thus no further change in the scaling. 
The  single dilaton intermediate state is $\ORD(F_D^2)$ which can be formed as the ratio of the first and the second 
entry in $g$ to yield $x^2 F^2$ and thus   $ d  \to d -y_\la - (d/2-1) = \De_{\Op}-(d/2-1)$ in that case. 
Finally, one gets  
\begin{equation}
\label{eq:TTm}
\CoE{\TEMT}_{\CDQCD} \propto    \frac{  c_{D} e^{-m_D x } }{(x^2)^{\De_\Op-(d/2-1)}} +
  \frac{1}{(x^2)^{\De_\Op}}( c_{2\pi} e^{- 2 m_D x}  +     c_{2D} e^{- 2 m_D x} )     \;,  
\end{equation}
where $c_D = \ORD(F_D^2)$ and $c_{2\pi,D}$ are $F_{\pi,D}$-independent as of above.  
The exponential behaviour follows from dominance of the lowest state in the specific channel and 
is well known from the study of Euclidean correlation functions in lattice applications. 
\end{itemize}
 Equating \eqref{eq:schematic2} with  \eqref{eq:TTm} one gets the result $\De_\Op = d-2$ 
 in accordance with \eqref{eq:DeOp}, serving as another consistency check.

\subsection{The single-soft theorem, $\matel{0}{\bar qq}{D}$ and  the dilaton decay constant  }
\label{app:single}

We may apply the  single  dilaton soft theorem  \eqref{eq:soft2}  to $\Opqq$ \eqref{eq:Oqq} which yields 
\begin{equation}
\label{eq:match2}
\matel{0}{\Opqq}{D} = -\frac{1}{F_D} \vev{ [Q_D,  \Opqq]} = - \frac{\Deqq}{F_D} \vev{\Opqq}   \;,
\end{equation}
as the remainder vanishes and we notice that the same result would follow from expanding \eqref{eq:qqmatch}
to linear order in the dilaton field. We may now make contact with \cite{LSD:2023uzj,LatticeStrongDynamics:2023bqp} 
where the following matrix element was defined
\begin{equation}
\matel{0}{\sum_q \mq \bar qq}{D} = m_D^2 \, F_S  \;,
\end{equation}
such that $F_S$ is RG invariant.  
Combining this equation together with  \eqref{eq:GMOR} and  \eqref{eq:match2} one gets 
\begin{equation}
\label{eq:FS}
F_S  = N_f \,  \frac{\Deqq}{(1+\gast)} \left( \frac{m_\pi^2 F_\pi^2}{m_D^2 F_D^2} \right) F_D \;.
\end{equation}
It has previously been obtained in \EQ 5 of \cite{LatKmi:2015non} from a soft-theorem  and from the EFT 
cf. \EQ 9 in \cite{LSD:2023uzj} 
($y = \Deqq $ in their notation).\footnote{A distinctive feature is that in  \cite{LSD:2023uzj}   
a generic potential \eqref{eq:V} is assumed on top of $\mq \neq 0$ with $\Deqq = y$ and this leads to 
$f_{\pi,D}$ and $F_{\pi,D}$ where the lower case quantities are the one in the chiral limit. 
Using that $F_\pi/F_D = f_\pi/f_D$  \cite{LSD:2023uzj}  we can escape these difficulties in principle 
and if we use the values in \cite{LatticeStrongDynamics:2023bqp}  we of get an agreement of the right- 
and left-hand within $7\%$ which is well within the errors. This is of no surprise as according to 
our understanding $F_S$ went into the fit in  \cite{LSD:2023uzj}. This is just a confirmation of their numbers.}   
We may use this formula to assess the zero chiral mass hypothesis ($\mbase =0$) as then the dilaton mass 
ought to be well approximated by  the dilaton GMOR-relation \eqref{eq:D-GMOR}. 
Using the latter in \eqref{eq:FS}  we get a remarkably simple relation  between $F_D$ and $F_S$ 
(with $\gast=1$)\footnote{The relation \eqref{eq:nice} is consistent with the large-$N_c$ considerations 
in \SEC\ref{sec:largeNc}. If  $F_D = \ORD(\Nc)$ then the same holds for $F_S = \ORD(\Nc)$  since  the dilaton does not couple to the closed $\bar qq$-correletor at leading order in $N_c$.}
\begin{equation}
\label{eq:nice}
 F_D |_{\mbase=0} =   2 F_S   \;,
\end{equation}
that is the coupling of the dlaton to the EMT-  and the $\bar qq$-operator 
respectively. We may then  define the quantity 
\begin{equation}
R_{S,\Nf} = \sqrt{\frac{\Nf}{2}} \frac{m_\pi F_\pi}{m_D F_S} \;, \quad R_{S,\Nf} \big|^{\LO}_{ \mbase=0} = 1 \;,
\end{equation}
which is unity in the  $\mbase \to 0$ limit at LO.  Using the LSD-data  \cite{LatticeStrongDynamics:2023bqp} one finds 
\begin{equation}
\label{eq:RS}
R_{S,8} \big|_{ \text{LSD}}\approx 1.18 \pm 5\% \;, 
\end{equation}
where we have added the lattice uncertainties in quadrature.  Note that
since $m_\pi/F_\pi \approx 4$  is a similar to the ratio to the kaon mass to the pion 
decay constant in QCD  NLO-corrections could easily amount to $30\%$ (e.g. $m_D^2/m_\rho^2 \approx 0.4$). 
Hence, the $20\%$-proximity to unity  in \eqref{eq:RS} is thus remarkable.  
Notice that for QCD with $m_\pi/F_\pi \approx 4$ the dilaton (or $\sig$-meson) is a stable bound state  \cite{Rodas:2023twk}, suggesting that the same is the case in these simulations. 
Moreover,  $\mbase \neq 0$ would lead to 
\begin{equation}
R_{S,\Nf} =  \frac{1}{\sqrt{1+x^2}}   \leq  1 \;,  \quad x \equiv \frac{\mbase}{(m_D)_{\text{GMOR}}} \;,
\end{equation}
where $(m_D)_{\text{GMOR}}$ is  the mass from the dilaton GMOR-relation \eqref{eq:D-GMOR}.
 We conclude that for $N_f =8$ 
the data do not exclude a massless dilaton in the chiral limit.\footnote{If we take $F_\pi^2/F_D^2 = 0.1089(41)$  \cite{LSD:2023uzj} and $F_\pi = 0.021677(40)$ \cite{LatticeStrongDynamics:2023bqp}    and get $F_S \approx 0.033 $ from  \eqref{eq:nice}  
which is $30\%$ off from $F_S = 0.0254(17)$ \cite{LatticeStrongDynamics:2023bqp}.  In our view 
the analysis above is preferable since it is independent of the potential.}
It would be interesting to extend this analysis to the other simulations such as the  $SU(3)$-case with $\Nf=4$  
\cite{LatKMI:2016xxi,LatticeStrongDynamics:2018hun} 
or the $SU(3)$ sextet-representation \cite{Fodor:2012ty,Fodor:2015vwa,Fodor:2016pls}. One might be hopeful that this will happen in the foreseeable 
future.

  \section{On Singlet-Octet Mixing}
\label{app:mix}

The aim of this appendix is to give some more interpretation with regards to 
the quark content of the $\sig$-meson.  Unlike for flavoured mesons this is not a well-defined question.  
There are many ways to think about it  as summarised in   the excellent  review \cite{Pelaez:2015qba}.  

An instructive starting point is the following ratio of rates \cite{PDG22}
\begin{equation}
\label{eq:rf0}
r_{f_0 \to KK/\pi\pi} = \frac{ \Ga( f_0(980) \to K^+K^-)}{ \Ga( f_0(980) \to \pi^+\pi^-)} = 0.69(32) \;. 
\end{equation}
If the $f_0(980)$ roughly equal parts of $u,d,s$-quarks then 
phase space would dictate  $r_{f_0 \to KK/\pi\pi} \ll 1$. The surprisingly large 
rate to kaons could be explained by the $f_0$ having a large mount of $\bar ss$-quarks versus 
$\bar qq$-quarks  ($q=u,d$).  It is indeed the commonly accepted view that $f_0(980)$ has 
a high strange quark content.

It is helpful to consider the octet and singlet in a $\bar qq$-  and a tetraquark-basis
 \begin{alignat}{4}
\label{eq:qq}
&   \state{S^{I=0}_1}_{\bar qq} &\;=\;& \frac{1}{\sqrt{6}}(\bar uu + \bar dd + \bar ss)   \;,   \quad 
& &  \state{S^{I=0}_1}_{\bar qq \bar qq} &\;=\;& \frac{1}{\sqrt{6}}(\bar ss \bar uu +  \bar ss\bar dd +  \bar uu \bar dd) \;,
 \nonumber \\[0.1cm]
&\state{S^{I=0}_8}_{\bar qq} &\;=\;& \frac{1}{\sqrt{6}}(\bar uu + \bar dd - 2 \bar ss)  \;,   \quad
& &   \state{S^{I=0}_8}_{\bar qq \bar qq} &\;=\;& \frac{1}{\sqrt{6}}(\bar ss \bar uu +  \bar ss\bar dd - 2 \bar uu \bar dd)   \;. 
\end{alignat}
The difference in role of the strange quark in the two bases is apparent. 
With the central value from the $SU(3)_F$-analysis \eqref{eq:Oller}, one finds the following quark compositions 
\begin{alignat}{6}
\label{eq:sig}
& \state{\sig}_{\bar qq}  &\;\approx\;&   0.67(\bar u u + \bar dd)  &\;+\;& 0.28 \bar ss \;, \quad  & & 
\state{\sig}_{\bar qq \bar qq}  &\;\approx\;&   0.67(\bar s s \bar u u + \bar s s \bar dd)  &\;+\;& 0.28 \bar uu \bar dd 
\;,
 \nonumber \\[0.1cm]
& \state{f_0}_{\bar qq}  &\;\approx\;&   0.20(\bar u u + \bar dd)   &\;-\;& 0.95 \bar ss \;, \quad  & & 
\state{f_0}_{\bar qq \bar qq}  &\;\approx\;&   0.20(\bar s s \bar u u + \bar s s \bar dd)   &\;-\;& 0.95 \bar uu \bar dd \;.
\end{alignat}
We see that  the  strange quark content in the $f_0(980)$  is enhanced in the $\bar qq$-states  
and suppressed for the tetraquarks.  Hence the $\bar qq$-state interpretation harmonises 
with the commonly accepted picture that the $f_0(980)$ has a large strange quark content.\footnote{In fact the angle where this is picture is extremised  is $\theta \approx 35.7^\circ$, also 
known as ideal mixing.  The angle, in the original proposal of Jaffe,  is $\theta \approx - 54.7^\circ$, 
where $\state{\sig}_{\bar qq\bar qq} = \bar u u \bar dd$.}
  Other analyses  finding support for 
the $\bar qq$-interpretation, which  is not the common view, are  for example \cite{Maltman:1999jn,Cherry:2001cj,Moussallam:2011zg}.

\bibliographystyle{utphys}
\bibliography{../Dil-refs.bib}

\end{document}